\documentclass[fleqn,usenatbib]{mnras}

\usepackage{newtxtext,newtxmath}
\usepackage{lipsum}
\usepackage[T1]{fontenc}

\DeclareRobustCommand{\VAN}[3]{#2}
\let\VANthebibliography\thebibliography
\def\thebibliography{\DeclareRobustCommand{\VAN}[3]{##3}\VANthebibliography}

\usepackage{graphicx}	% Including figure files
\usepackage{amsmath}	% Advanced maths commands
\newcommand{\redd}{R$_{\rm Edd}$}

\title[QSO optical variability with M and REdd]{Optical variability in Quasars: Scalings with black hole mass and Eddington ratio depend on the observed timescales}
\author[P. Ar\'evalo et al.]{P. Ar\'evalo,$^{1,2}$\thanks{E-mail: patricia.arevalo@uv.cl)}
P. Lira,$^{3,2}$
P. S\'anchez-S\'aez,$^{4,5}$
P. Patel,$^{3,2}$
E. L\'opez-Navas,$^{1,2}$
E. Churazov$^{6}$ and
\newauthor L. Hern\'andez-Garc\'ia$^{1,5}$
\\
$^{1}$Instituto de F\'isica y Astronom\'ia, Universidad de Valpara\'iso, Gran Breta\~na 1111, Valpara\'iso, Chile\\
$^{2}$Millennium Nucleus on Transversal Research and Technology to Explore Supermassive Black Holes (TITANS)\\
$^{3}$ Departamento de Astronom\'ia, Universidad de Chile, Casilla 36D, Santiago, Chile\\
$^{4}$ European Southern Observatory, Karl-Schwarzschild-Str. 2, 85748, Garching, Germany\\
$^{5}$ Millennium Institute of Astrophysics (MAS),Monse\~nor S\'otero Sanz 100, Providencia, Santiago, Chile\\
$^{6}$ Max-Planck-Institut f\"ur Astrophysik, Karl-Schwarzschild-Str. 1, 85748, Garching, Germany}

\date{Accepted XXXX. Received YYYYYY; in original form ZZZZZ}

\pubyear{2023}

\begin{document}
\label{firstpage}
\pagerange{\pageref{firstpage}--\pageref{lastpage}}
\maketitle

\begin{abstract}
Quasars emission is highly variable, and this variability gives us clues to understand the accretion process onto supermassive black holes. We can expect variability properties to correlate with the main physical properties of the accreting black hole, i.e., its mass and accretion rate. It has been established that the relative amplitude of variability anti-correlates with the accretion rate. The dependence of the variance on black hole mass has remained elusive, and contradicting results, including positive, negative, or no correlation, have been reported. In this work, we show that the key to these contradictions lies in the timescales of variability studied (e.g., the length of the light curves available).By isolating the variance on different timescales in well-defined mass and accretion rate bins we show that there is indeed a \emph{negative} correlation between black hole mass and variance and that this anti-correlation is stronger for shorter timescale fluctuations. The behavior can be explained in terms of a universal variability power spectrum for all quasars, resembling a broken power law where the variance is constant at low temporal frequencies and then drops continuously for frequencies higher than a characteristic (break) frequency $f_b$, where $f_b$ correlates with the black hole mass.  Furthermore, to explain all the variance results presented here, not only the normalization of this power spectrum must anti-correlate with the accretion rate, but also the \emph{shape} of the power spectra at short timescales must depend on this parameter as well. 
\end{abstract}


\section{Introduction}
Establishing a statistically significant correlation between the observed brightness variability and the physical properties of quasars, such as the mass of the supermassive black hole (M) and accretion rate normalized to the Eddington limit (\redd), is the goal of many current studies. With the arrival of new and revolutionary large-scale surveys in time-domain astronomy, these potential correlations could allow the characterization of millions of supermassive black holes, which has been unfeasible until now. Early works found a clear anti-correlation between luminosity and variance in the brightness fluctuations of quasars \citep{Angione72,Hook94,Cristiani97,VandenBerk04}. However, as luminosity is the product of the black hole mass and normalized accretion rate (\redd), the dependence of variability on these more intrinsic properties remained hidden. More recently, a statistically significant anti-correlation between variance and accretion rate has emerged \citep{Kelly09,MacLeod10,Zuo12,Kelly13,Simm16,Rakshit17,Sanchez-Saez18,Lu19}, but whether there is a correlation with black hole mass has remained a matter of debate. Positive correlations were found in \citet{Wold07,Wilhite08,MacLeod10,Lu19}, while negative correlations were found by \citet{Kelly09,Kelly13} and no or unclear correlations were reported by \citet{Zuo12,Simm16,Rakshit17} and \citet{Li18}. These conflicting results can be reconciled when a well-defined sample of quasars is analyzed considering the different timescales of variation.

Our first step in this study was to constrain our sample to a narrow redshift range so that all analyses are performed at the same rest-frame wavelength, and the same intrinsic emission of the quasar is captured. Secondly, all quasars have homogeneous estimations of their physical properties, mass (M), and accretion rate normalized to the Eddington limit (\redd ). Third, the variance analysis was done by isolating the variations on different timescales, ranging from 30 to 300 days in the quasar rest-frame, and the variability analysis was conducted separately for each of them. This method is similar to measuring the variance with light curves of different lengths, which can only capture variations on timescales shorter than the length of the light curve, and with different binnings, which can only capture variations on timescales longer than the width of the time bins. Fourth, we sampled carefully defined bins in M {\em and\/} \redd\ so that both properties are disentangled, and their correlation with the variance determined at different timescales could be assessed separately.

\section{Sample selection}\label{sec2}
We selected quasars with optical spectral classification from the catalog of \citet{Rakshit20}. These authors performed a homogeneous analysis of all quasar spectra observed by the SDSS and reported, among other many quantities, black hole masses and Eddington ratios for the majority of their half-million sources. The Eddington ratio \redd\ has been estimated by taking the ratio of $L_{\rm bol}$ to Eddington luminosity $L_{\rm Edd} = 1.3 \times 10^{38}(M_{\rm BH}/M_\odot)$ erg s$^{-1}$ to measure the accretion rate.

The amplitude of variability depends on the rest-frame wavelength of the emission that is captured by the light curves used \citep[e.g.][]{Sanchez-Saez18}. If a sample contains quasars at different redshifts, this dependence needs to be accounted for before searching for the dependence of the variance on other parameters. In order to minimize the effects of different rest-frame wavelengths, we selected only quasars for a narrow redshift bin $z=0.6-0.7$. The median redshift is high enough to include a large number of sources and low enough to allow the H$\beta$ line to fall well within the SDSS spectral range. With this requirement, all selected sources have H$\beta$-derived masses, which is the best-calibrated single-epoch mass estimator. Further requirements include (i) a reported (statistical) error on the black hole mass less than 0.2 dex, which allows us to measure variability amplitudes as a function of mass and accretion rate for fine mass bins, and (ii) a g-band magnitude $g<20.5$ in the SDSS data release 12 photometric catalog. This selection returned 5881 objects. Other spectral considerations regarding the effect of emission lines in the observed range are discussed in Appendix \ref{ap:spectral}.

Optical light curves were extracted for these 5881 selected quasars from the Zwicky Transient Facility \citep[ZTF][]{Masci19} Data Release 14, obtaining data for 5651 objects. These light curves cover the period March 2018 to September 2022 and have approximately 4-day cadence, with yearly gaps, although some regions of the sky have been observed much more frequently. In order to homogenize the light curves of different objects, we require observations to be taken at least 1 day apart, retaining only the first observation in a given night and discarding the rest. This procedure produces a slightly more homogeneous sample of light
curves for different objects (in terms of cadence), as
most objects only have one observation in a given night, while a few
have many observations in a few nights. We chose light curves in the $g$ band because the variations are stronger here than in the other available band ($r$), as expected from the anti-correlation between variance and restframe emitted wavelength seen in quasars \citep[e.g.][]{Sanchez-Saez18}. The $g$ band is also less contaminated by the star light of the host galaxies. 

The photometric quality of the individual observations was controlled using the \emph{limitmag} values present in the DR14 light curves. We retained only observations with \emph{limitmag}$\ >20$, which ensures that objects with $g \leq 20$ could be detected at least at the 5$-\sigma$ level in all epochs. We note that \emph{limitmag} is a property of the observation, not of the object, so this process only removes observing nights with bad conditions, regardless of the brightness of each object. This filtering removed 5--10\% of the epochs in each light curve. We also selected only observations with processing quality flag \emph{catflags}$\ =0$. The ZTF light curves are composed of observations performed on different CCDs of the detector, which might have cross calibration offsets and in most light curves the different CCDs are approximately alternated. We circumvented the cross calibration uncertainties by constructing light curves for individual CCDs. This procedure results in multiple light curves for some objects.  We calculated the average flux of the filtered ZTF light curves and removed objects whose corresponding $g$-band magnitude was greater than 20. We further restricted the sample to include only light curves at least 900 days long in the observer frame and with at least 90 data points in the single-CCD, quality-filtered data sets. This sampling allows us to measure fluctuations on timescales of several tens to hundreds of days, which are similar to previous quasar variability studies (see references in the Introduction).

 The final sample contains 4770 individual objects; for 616 objects there were two acceptable light curves; for 21 objects there were three or more light curves, resulting in 5433 valid light curves. Multiple lightcurves for single sources were kept for the analysis considering that they represent a different realization of the same variability process, with a different noise pattern. Therefore including both in the calculation of median values results in better estimates of the variance. The average flux of each light curve was subtracted and the resulting zero-mean light curves were further normalized by their respective (pre-subtraction) mean so that the amplitude of variations and the variance can be directly compared between objects of different flux levels. The valid light curves have a mean (median) length in the observer frame of 1544 (1547) days with a standard deviation of 60 days and contain a mean (median) of 245 (233) good data points with a standard deviation of 90 data points. 

\section{Estimation of the variance}

We isolated variations on different timescales using the Mexican Hat filter \citep{Arevalo12}, which is ideally suited to deal with uneven sampling light curves with gaps. In short, the light curves are  convolved with two Gaussian kernels of comparable widths and takes the difference of the convolved light curves, correcting for effects of the sampling pattern, and calculates the variance of the filtered light curve. For a given value of Gaussian width $\sigma$, the filter applied on the power spectrum peaks at $k_p =0.225/\sigma$ and has a  width of $1.16 k_p$. This relatively broad filter in frequency space has the net effect of averaging together the power of a few independent frequencies, reducing the scatter in power from independent frequency bins, expected from red noise variability processes. This reduction comes at the cost of limiting the spectral resolution. The filtered powers are estimates of the normalized power density and have units of days. We estimated the observational noise contribution to the filtered power from the reported errors on the flux, as described below, and subtracted this value from the measured power before plotting and fitting. Finally, the normalized power estimates were converted into dimensionless variance estimates by multiplying each one by the peak frequency $k_p$ of each frequency filter.

We stress that the variability timescales studied here do not
necessarily correspond to characteristic timescales of the quasars, they are simply a few selected timescales at which we can reliably estimate a band-limited variance. The filtering process described above is akin to cutting the light curves in segments slightly longer than the timescale studied, therefore removing fluctuations on longer timescales, and binning and averaging the data points on bins slightly shorter than the selected timescale, thereby removing faster variations. The published correlations between variance and mass cited above use either the total variance of the light curves, which is normally dominated by the variations on the longest timescales available (i.e. defined by the length of the light curves), or an estimate of the variance at a given timescale, for example by evaluating the structure function \citep[e.g.][]{deVries2005} at a single value of the time delay $\tau$. The present analysis studies four separate timescales, which can be interpreted as studying the correlations we would find if the objects had been observed by four monitoring campaigns of different lengths and samplings, or if the structure function were evaluated at four different values of $\tau$.

The contribution of observational noise to the filtered variance was estimated through simulated light curves with additional noise as described in Appendix \ref{ap:noise}.

\section{Results}
\begin{figure*}%
\centering
\includegraphics[width=0.49\textwidth]{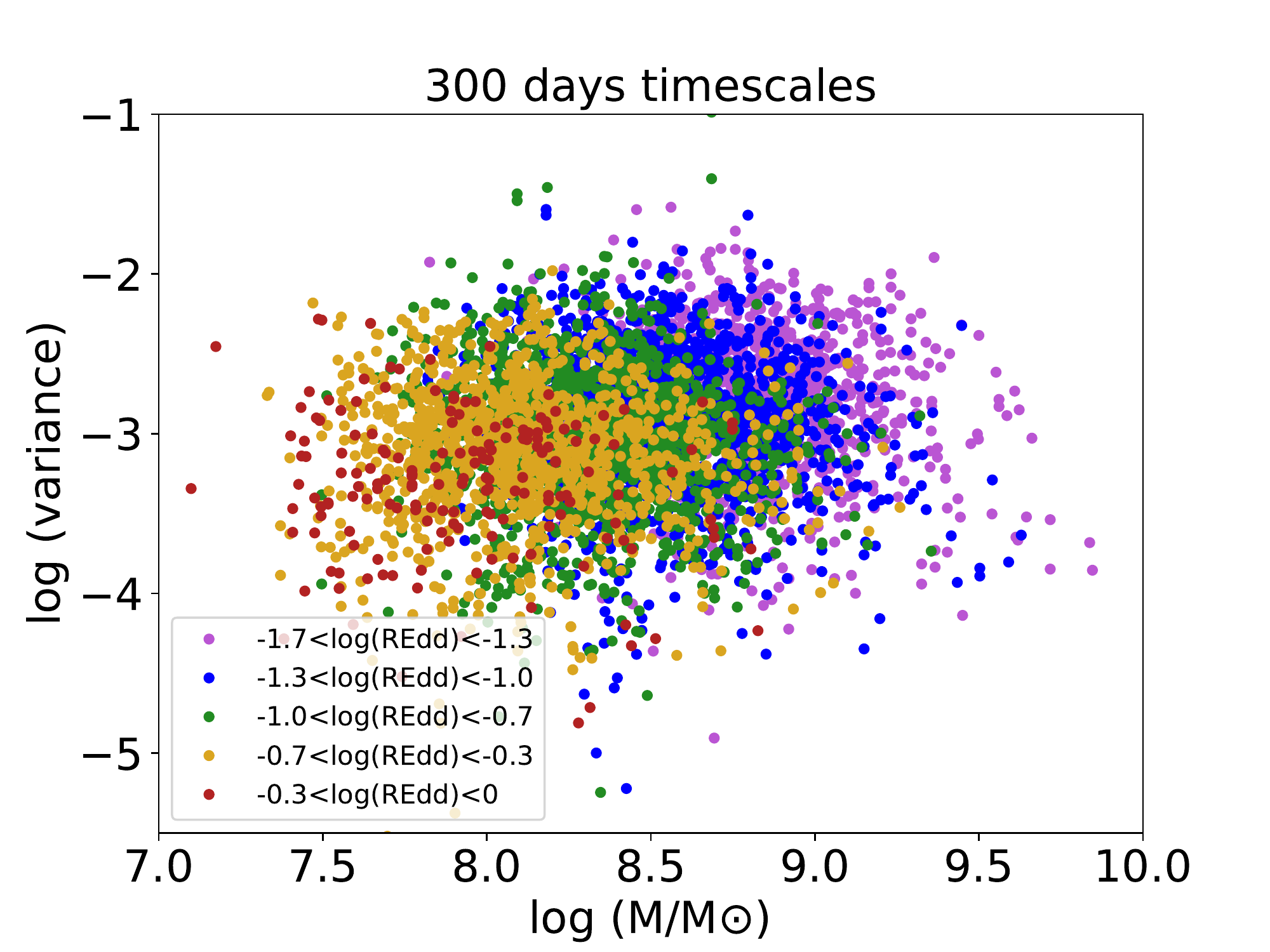}
\includegraphics[width=0.49\textwidth]{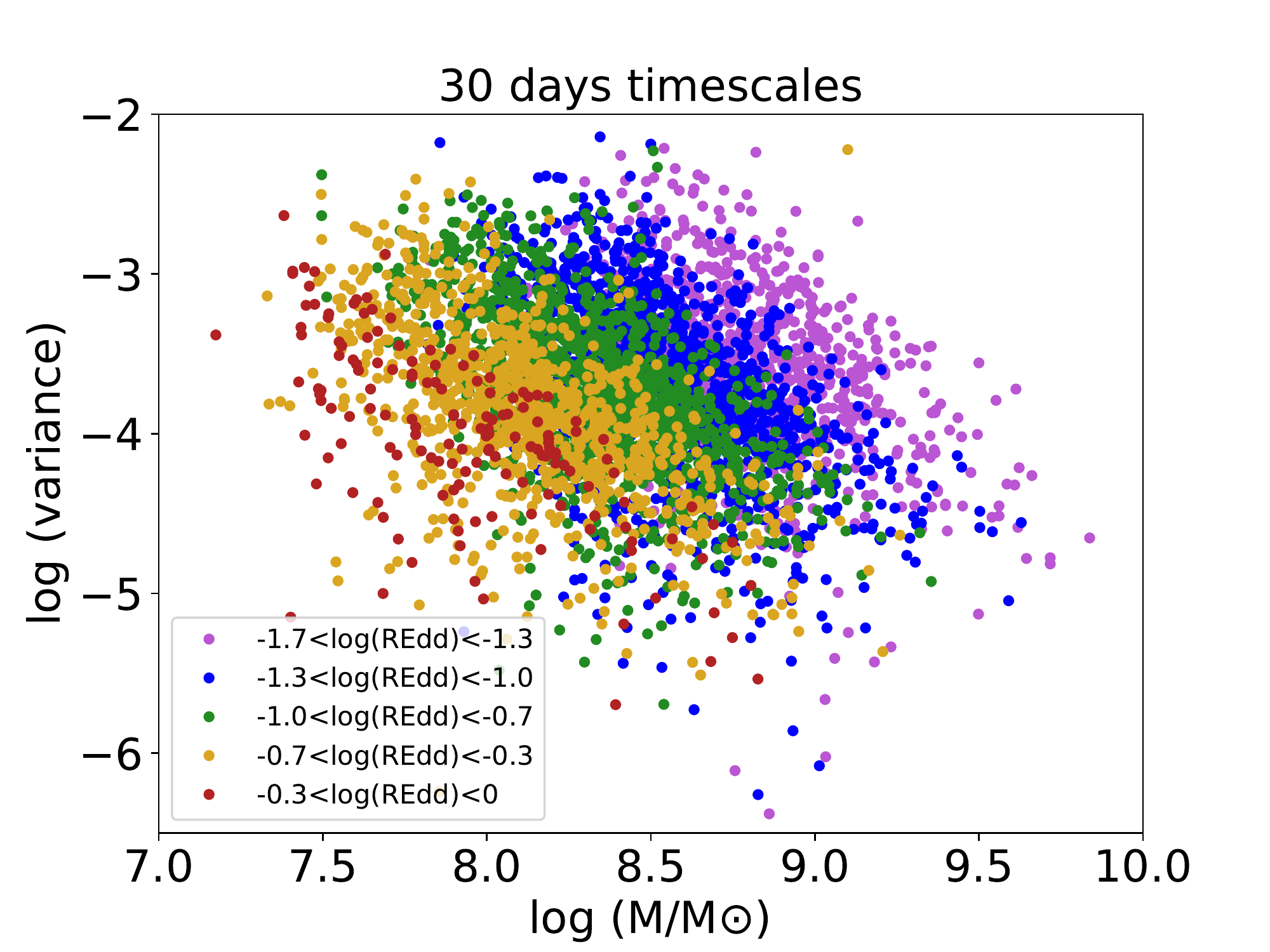}

\caption{ Variance as a function of black hole mass (M): the panels show this relation for the same set of quasars but for the variance associated with two different timescales, i.e. 300 days and 30 days for the left and right panels, respectively. Circles show individual variance measurements, with colors assigned by ranges of \redd\ as noted in the legend. At the longest timescales, the sample selection effects (anti-correlation between M and \redd ) and the known anti-correlation
between \redd\ and variance lead to the weak spurious correlation between variance and M. At shorter timescales the anticorrelation
between the variance and mass is evident. The shift between the groups of points with different colors, i.e. different \redd , reflects the known anti-correlation between \redd\ and variance.
}\label{scatterplots}

\end{figure*}

The timescales selected for the detailed study satisfy two criteria: they are covered at least twice in the length of the light curve and are sufficiently separated in frequency space to produce independent variance estimates, considering the width of the Mexican filter $\delta k\sim k$. For $z = 0.65$, this results in a longest timescale of about $T=1500/2/(1+z)=454$ days in the rest-frame of the quasar. To be conservative, we limited the longest studied timescale to 300 days in the quasars rest-frame and chose a separation by a factor $\sim 2$ for the other timescales: 150, 75, and 30 days.
 The variance as a function of black hole mass is plotted in Fig. \ref{scatterplots} for two of the four different timescales probed--- left: 300 days, right: 30 days. Variations on timescales of 150 and 75 days show an intermediate behavior. All timescales refer to the quasar rest-frame, i.e., T$_{\rm rest}$=T$_{\rm obs}/(1+z)$.

To explore the dependence of variability on quasar parameters, we split the sample according to their mass and accretion rate. We grouped the quasars in bins of width 0.33 dex in both M and \redd, starting from log(\redd) = -2 and from log(M/M$\odot$) = 7.5. For each M-\redd\ bin, we calculated the median variance, median M, and median \redd. The standard errors on these medians were estimated using the bootstrapping method, calculating the standard deviation of the medians of 1000 random re-samples for each M-\redd\ bin. The variance median was calculated as the median value of the median variances of these bootstrapping samples. We discarded bins with less than 10 quasars for all the analyses described below. 

\begin{figure*}%
\centering
\includegraphics[width=0.49\textwidth]{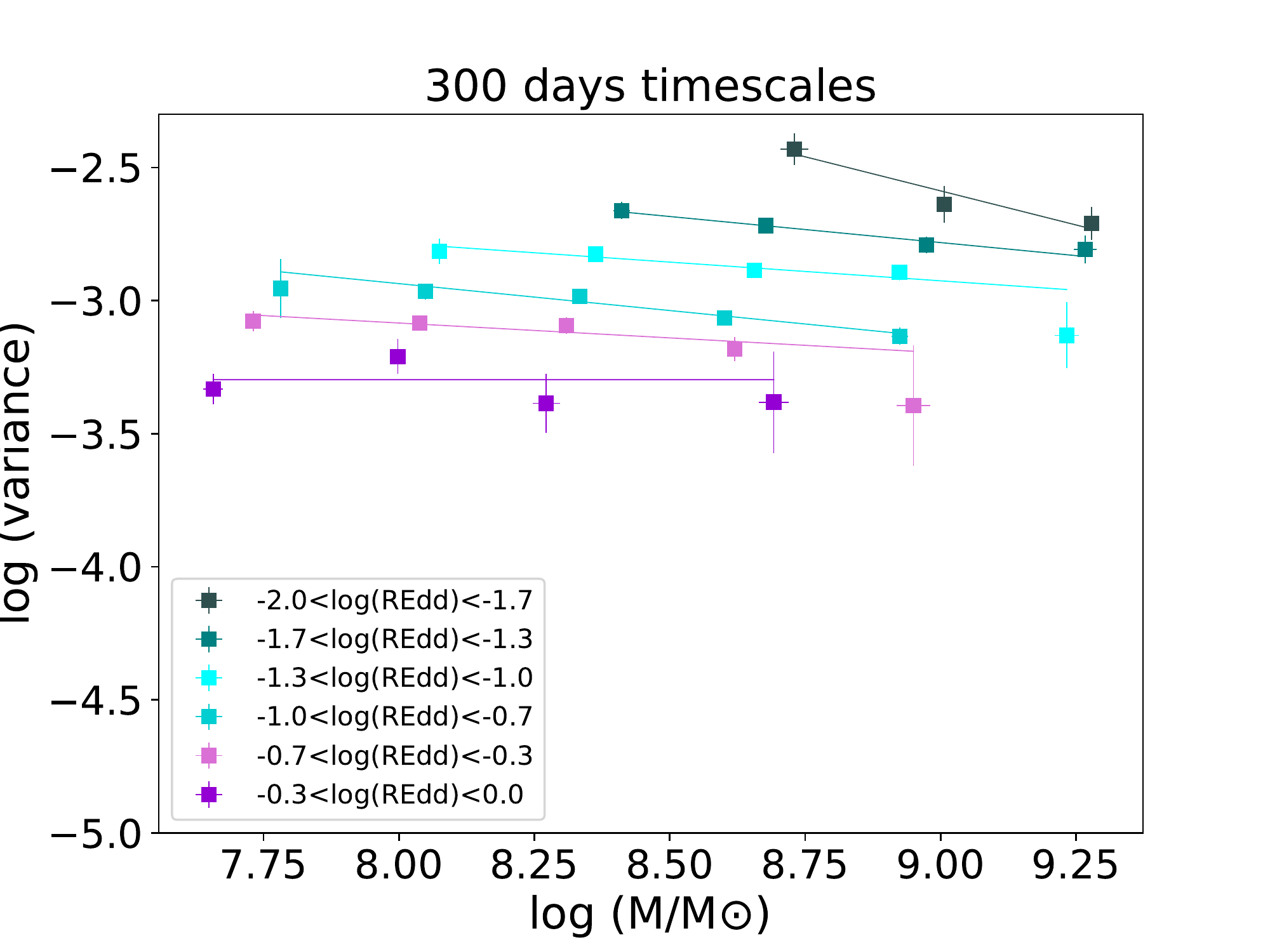}
\includegraphics[width=0.49\textwidth]{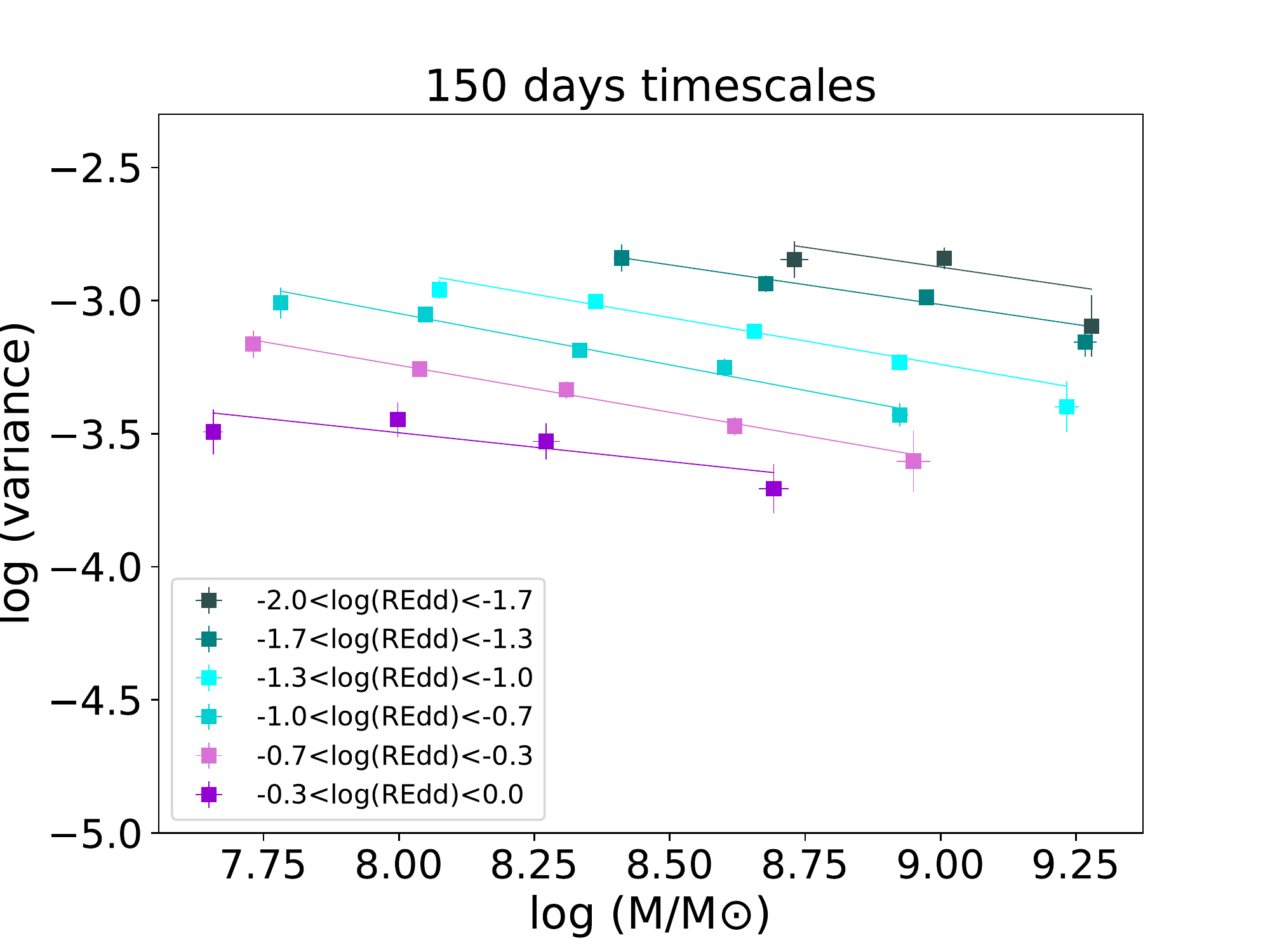}
\includegraphics[width=0.49\textwidth]{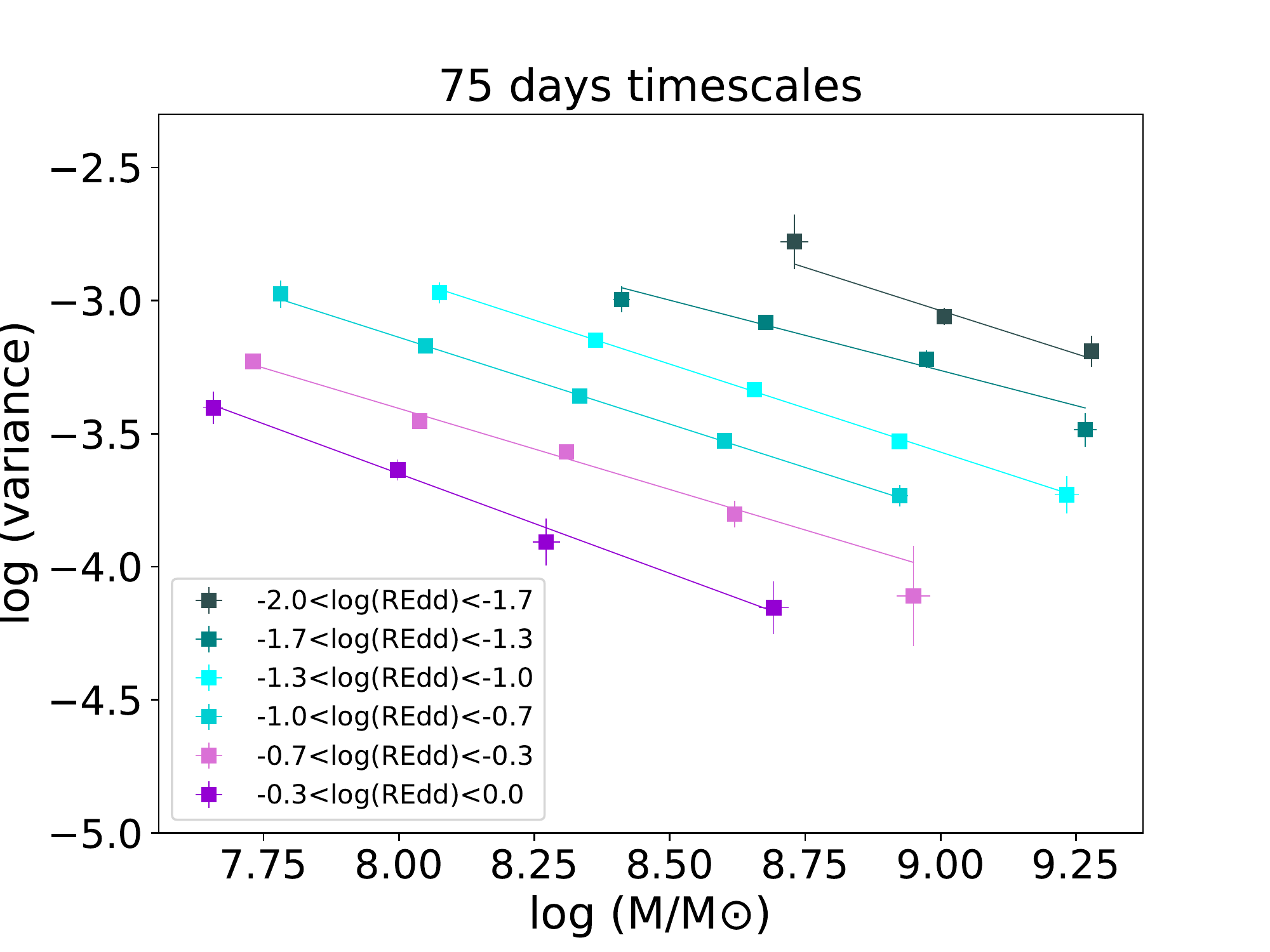}
\includegraphics[width=0.49\textwidth]{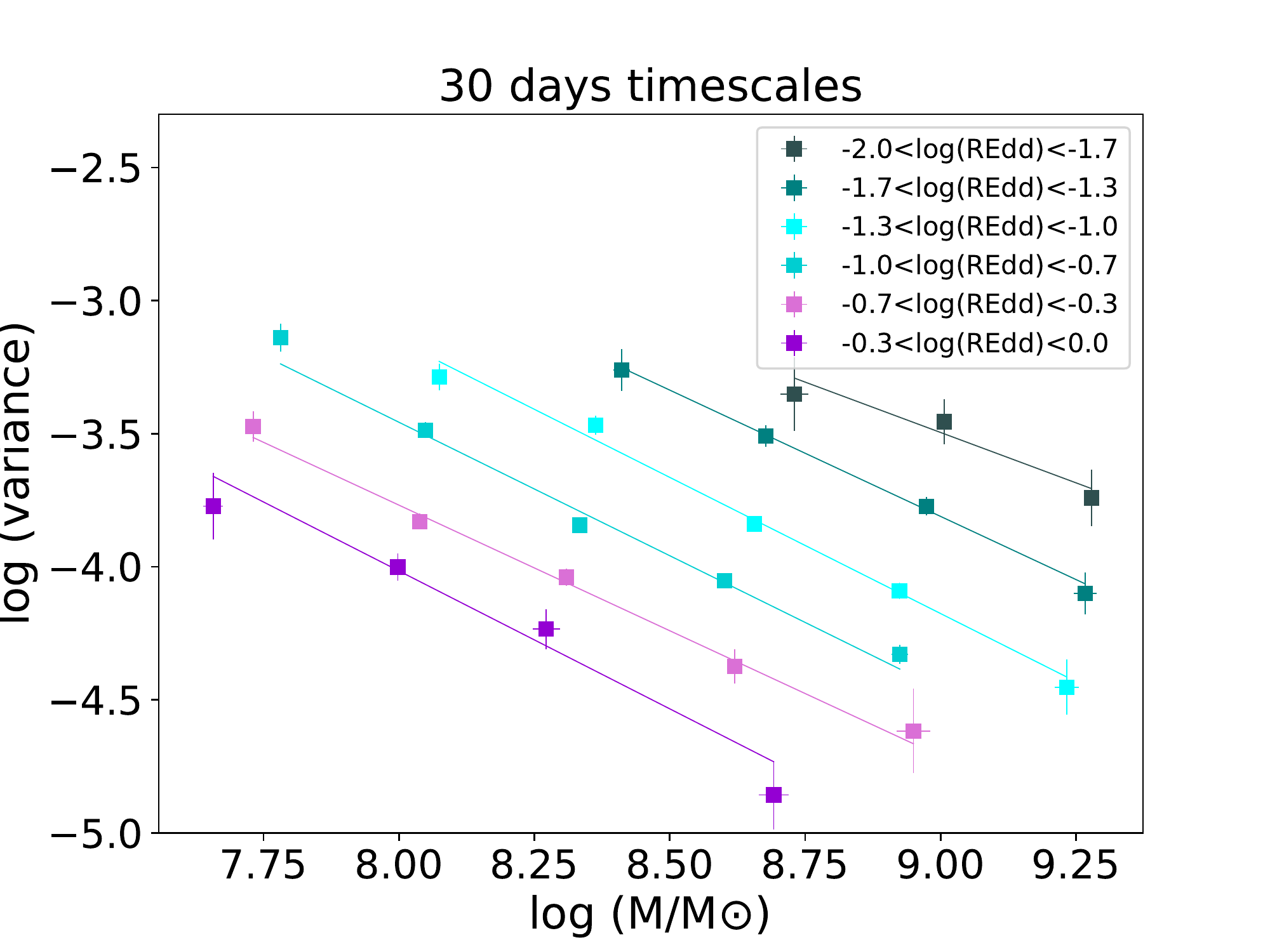}
\caption{Variance as a function of black hole mass, for sub-samples restricted to narrow ranges in accretion rate, as labeled in the plots, for variations on four different timescales, top left to bottom right: 300, 150, 75, and 30 days. The dependence of variance on M is very weak for the long term fluctuations but is clear for variations on timescales of 150 days and becomes stronger for shorter timescale variations. For all timescales, the variance is larger for lower \redd, which produces the large spread in variance at a given mass for the sample as a whole. All the error bars on the median variance of the binned data (markers) were calculated as the root-mean-squared scatter of the medians obtained by bootstrapping using 1000 re-samples per M-\redd\ bin.}\label{variance_vs_mass}
\end{figure*}

\begin{figure*}%
\centering
\includegraphics[width=0.49\textwidth]{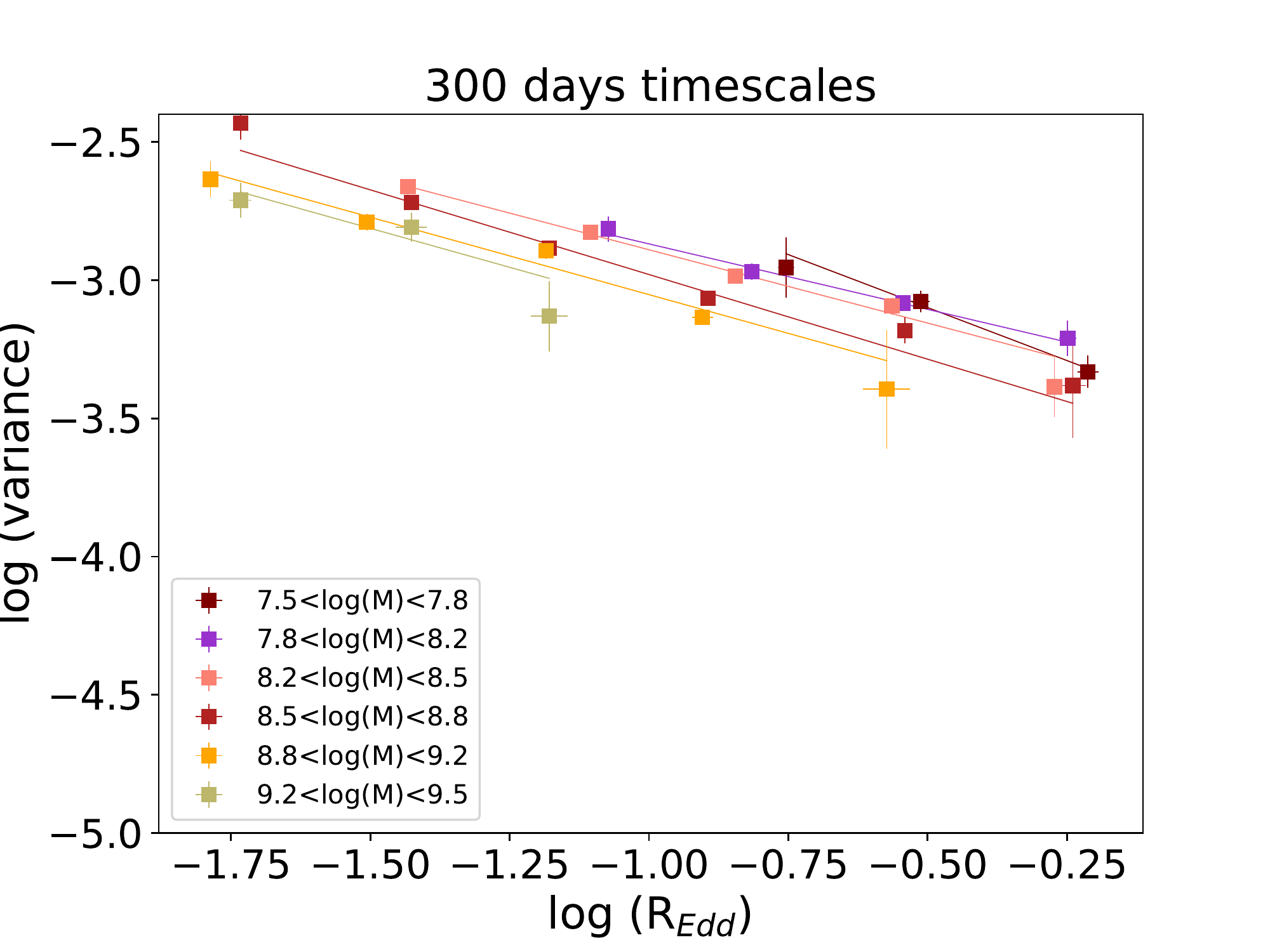}
\includegraphics[width=0.49\textwidth]{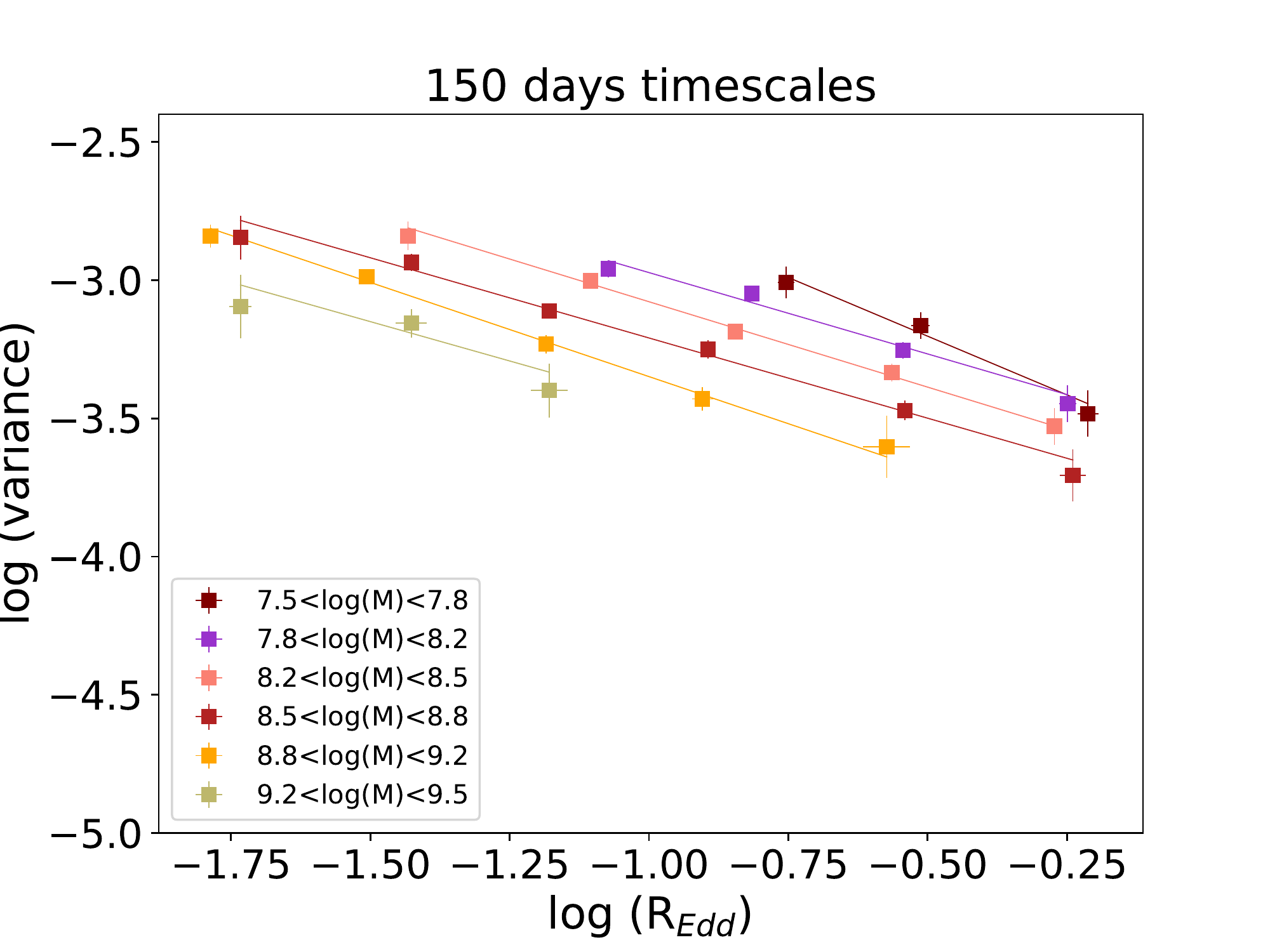}
\includegraphics[width=0.49\textwidth]{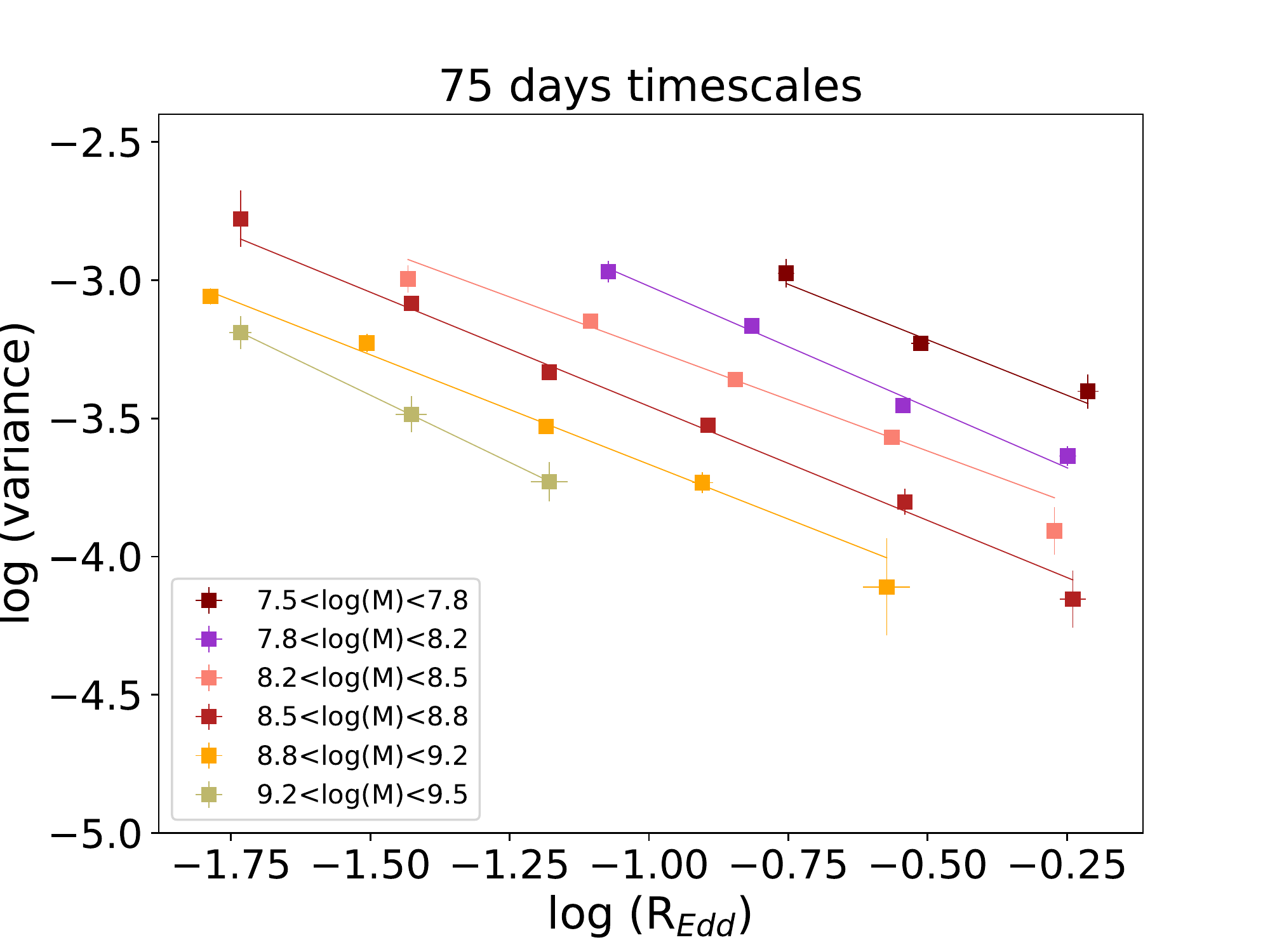}
\includegraphics[width=0.49\textwidth]{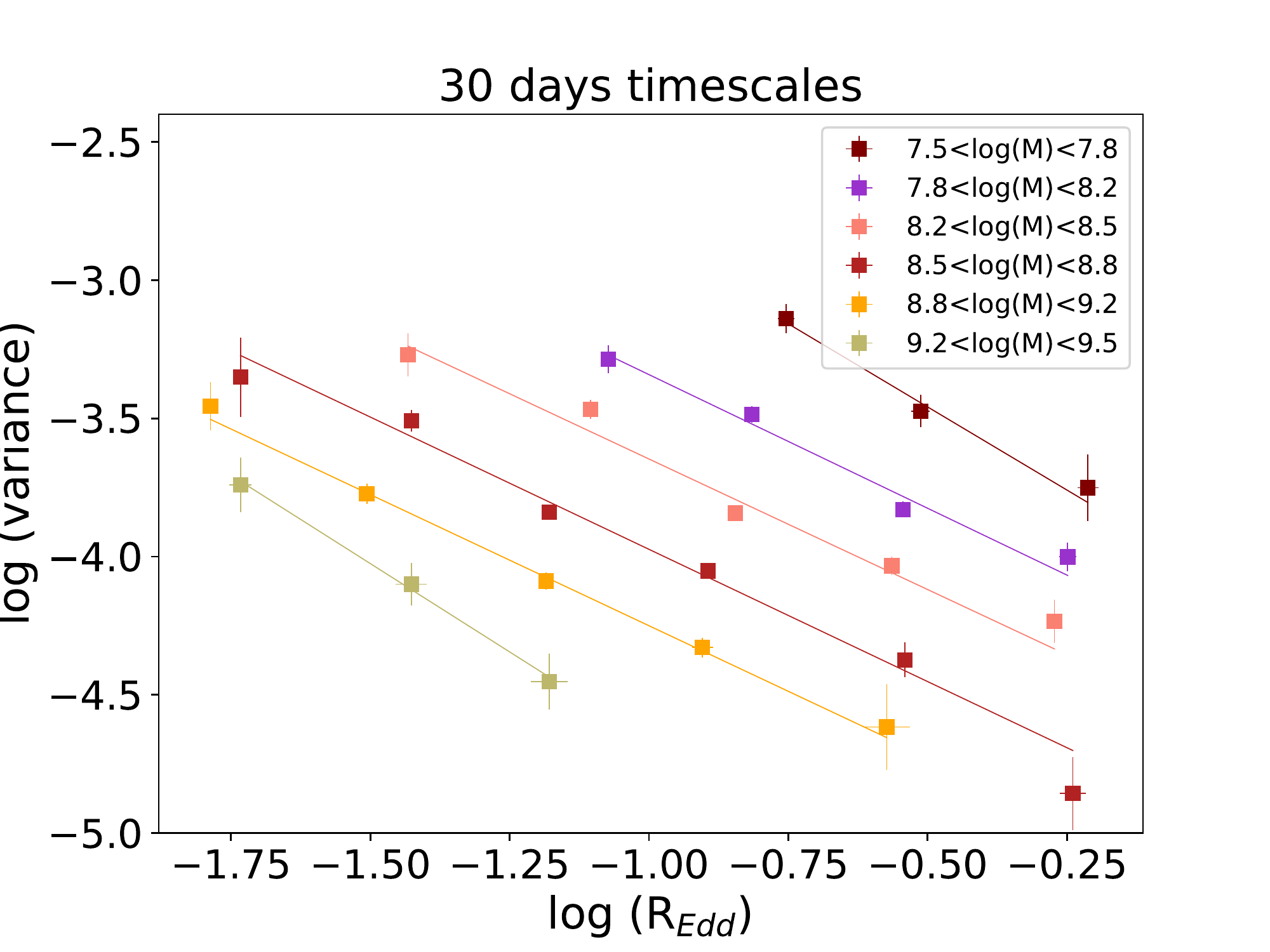}
%\includegraphics[width=0.5\textwidth]{var3_REdd_final.pdf}
%\vspace{0.5cm}
\caption{Variance as a function of \redd, for sub-samples restricted to narrow ranges in mass, as labeled in the plots, for variations on four different timescales, top left to bottom right: 300, 150, 75, and 30 days. The dependence of variance on \redd\ is significant on all variability timescales. The separation between mass bins increases towards shorter timescales, as expected from Fig.\ref{variance_vs_mass}. Interestingly, the slope of the relationship between variance and \redd\ changes with timescale, becoming steeper for shorter timescales.}\label{variance_vs_REdd}
\end{figure*}

These median variances are plotted as a function of black hole mass in Fig. \ref{variance_vs_mass} color-coded by the median \redd\ of each bin, and as a function of \redd in Fig.\ref{variance_vs_REdd}, color-coded by the median M of each bin. The different panels correspond to the four different timescales of variability studied.

\subsection{Correlations of variance with M, \redd\ and Bolometric Luminosity}
To evaluate the significance of the correlations between \emph{un-binned} variance and black hole mass M, we calculated the Spearman rank coefficients ($\rho$), which vary between -1 and 1 with 0 implying no correlation, and their $p$-values, which roughly indicate the probability of an uncorrelated system producing data sets that have the same Spearman correlation coefficient. A Spearman correlation coefficient $\rho$ = -1 implies an exact monotonic negative relationship. This analysis was repeated for sub-samples constrained to narrower ranges in \redd\ in order to separate the dependence of variance on mass from the dependence of variance on accretion rate. The ranges in \redd\ with sufficient quasars and the resulting values of $\rho$ and $p$ are summarized in Table \ref{spearman}. The remaining quasars are in ranges of  \redd\ with too few points to make a significant correlation analysis. 

\begin{table*}
\centering
\begin{tabular}{|c|cccc|c|}
&\multicolumn{4}{c}{Spearman rank coefficient (p-value) between variance and M }\\
\hline
  &  300 days & 150 days  & 75 days &30 days & N\\
  \hline
  \hline
Full sample &0.15 (1.4e-29) &0.04(1e-3) &-0.07(5e-7) & -0.14(2e-26)& 5433\\
\hline
-2.0$<$ log(\redd)$<$ -1.7 &-0.19(0.007)&-0.20(4e-3)&-0.35(2e-7) &-0.24 (6e-4) & 210\\
-1.7$<$ log(\redd)$<$ -1.3 &-0.09(0.006)&-0.18(9e-9)&-0.35(1e-30) &-0.33 (4e-27) & 1018\\
-1.3$<$ log(\redd)$<$ -1.0 &-0.10(1e-4)&-0.23(1e-18)&-0.41(7e-62) &-0.36(1e-45) &1462\\
-1.0$<$ log(\redd)$<$ -0.7  &-0.08(1e-3)&-0.23(1e-18)&-0.41(6e-61)& -0.36(8e-45)&1462\\
-0.7$<$ log(\redd)$<$ -0.3  &-0.04(0.17)&-0.24(1e-15)&-0.42(6e-47) & -0.38(1e-38)&1062\\
-0.3$<$ log(\redd)$<$ 0.0  &-0.04(0.59)&-0.29(1e-4)&-0.58(5e-17)& -0.50(2e-12)&174\\
\hline
     \end{tabular}

     \caption{Spearman rank coefficient and associated p-value between the variance and M, for variances measured at different variability timescales. In the full sample, the Spearman coefficients are close to zero (i.e., almost no correlation) but due to the large number of points, the largest of these correlations are significant. Importantly, the correlation coefficients are positive for long timescales and negative for short timescales. Separating the data according to their log(\redd) results in \emph{negative} correlation coefficients for all timescales, with stronger and more significant correlations at shorter timescales. The last column shows the number of light curves included in each range in \redd.} 
     \label{spearman}
 \end{table*}
 
To estimate the relation between variance and mass independently of the \redd, we fitted the logarithm of the median variance of each M--\redd\ bin described above, assuming a linear form $\log({\rm var})=a\times \log{(\rm M/M_8})+b$ using Orthogonal Distance Regression (ODR algorithm implemented in SciPy) and the errors on both axes, separately for each range in \redd. The best fitting values of the parameters $a$ and $b$ and their 1$-\sigma$ errors are tabulated in Table \ref{tab:var_vs_mass}. The relation between variance and \redd\ was similarly modeled separately for each range in M as $\log({\rm var})=a\ \times$ log(\redd/0.1) $+\ b$. The results of these fits are tabulated in Table \ref{tab:var_vs_REdd}. These best-fitting linear relations to the binned data are over-plotted to the data points in Fig.~\ref{variance_vs_REdd}.

 \begin{table}
     \centering
     \begin{tabular}{r r | r r r r}
log(\redd)& timescale[d] & \multicolumn{2}{c}{$a$} & \multicolumn{2}{c}{$b$}  \\
\hline
\hline
-1.7 -- -1.3
&300 & -0.19&$\pm{0.03}$ & -2.68&$\pm{0.01}$\\
&150 & -0.30&$\pm{0.08}$ & -2.86&$\pm{0.07}$\\
&75 & -0.53&$\pm{0.09}$ & -3.00&$\pm{0.03}$\\
&30 &-0.93  &$\pm{0.05}$ & -3.34&$\pm{0.02}$\\
\hline
-1.3 -- -1
&300 &-0.13&$\pm{0.05}$ & -2.85&$\pm{0.01}$\\
&150 & -0.35&$\pm{0.05}$ & -3.06&$\pm{0.01}$\\
&75 & -0.66&$\pm{0.02}$ & -3.24&$\pm{0.01}$\\
&30 & -1.03 &$\pm{0.04}$ &-3.66 &$\pm{0.02}$\\
\hline
-1 -- -0.7 
&300 &-0.20 &$\pm{0.03}$ &-3.04 &$\pm{0.01}$\\
&150 & -0.38&$\pm{0.04}$ & -3.24&$\pm{0.01}$\\
&75 & -0.65&$\pm{0.01}$ &-3.46&$\pm{0.01}$\\
&30 & -1.01 &$\pm{0.09}$ & -3.96&$\pm{0.03}$\\
\hline
-0.7 -- -0.3
&300 & -0.11&$\pm{0.05}$ & -3.14&$\pm{0.02}$\\
&150 & -0.35&$\pm{0.02}$ & -3.42&$\pm{0.01}$\\
&75 & -0.65&$\pm{0.04}$ & -3.46&$\pm{0.01}$\\
&30 & -0.94 &$\pm{0.06}$ & -4.24&$\pm{0.03}$\\
\hline
-0.3 -- 0
&300 & 0.00&$\pm{0.17}$ & -3.30&$\pm{0.11}$\\
&150 & 0.20&$\pm{0.11}$ & -3.60&$\pm{0.06}$\\
&75 & -0.61&$\pm{0.05}$ & -3.71&$\pm{0.01}$\\
&30 &  -1.04&$\pm{0.17}$ & -4.53&$\pm{0.08}$\\
\end{tabular}
\caption{Results of a linear fit to the relation between log(variance) and log(M), for the four  timescales shown in in Fig. \ref{variance_vs_mass} and the five ranges in \redd\ with four or more  bins in black hole mass. The parameter $a$ represents the slope of the relation and parameter $b$ the  $\log$(variance) at $\log($M/M$_\odot)=8.5$, i.e., $\log$(variance)= $a\times\log$(M/M$_{\rm 8.5}$) + b. For each bin in \redd , the slope $a$ becomes more negative when the variance is measured for shorter timescale fluctuations. }
     \label{tab:var_vs_mass}
 \end{table}

 \begin{table}
     \centering
     \begin{tabular}{r r | r r r r}
log(M)& timescale[d] & \multicolumn{2}{c}{$a$} & \multicolumn{2}{c}{$b$}  \\
\hline
\hline
7.8 -- 8.2
&300 &-0.47 &$\pm{0.04}$ & -2.87&$\pm{0.02}$\\
&150 & -0.60&$\pm{0.08}$ &-2.97 &$\pm{0.03}$\\
&75 & -0.88&$\pm{0.08}$ &-3.02 &$\pm{0.03}$\\
&30 & -0.97&$\pm{0.13 }$ & -3.34&$\pm{0.05}$\\
\hline
8.2 -- 8.5
&300 & -0.53&$\pm{0.04}$ &-2.89 &$\pm{0.01}$\\
&150 &-0.62 &$\pm{0.03}$ & -3.08&$\pm{0.01}$\\
&75 &-0.74 &$\pm{0.06}$ & -3.25&$\pm{0.02}$\\
&30 & -0.95&$\pm{0.14 }$ & -3.64&$\pm{0.04}$\\
\hline
8.5 -- 8.8 
&300 & -0.61&$\pm{0.06}$ &-2.98 &$\pm{0.02}$\\
&150 &-0.58 &$\pm{0.03}$ & -3.21&$\pm{0.01}$\\
&75 &-0.82 &$\pm{0.05}$ & -3.46&$\pm{0.01}$\\
&30 & -0.96&$\pm{0.08 }$ & -3.97&$\pm{0.02}$\\
\hline
8.8 -- 9.2
&300 & -0.55&$\pm{0.08}$ & -3.05&$\pm{0.03}$\\
&150 & -0.68&$\pm{0.03}$ & -3.35&$\pm{0.01}$\\
&75 &-0.79 &$\pm{0.04}$ &-3.67 &$\pm{0.02}$\\
&30 & -0.95&$\pm{0.04 }$ & -4.25&$\pm{0.01}$\\
\hline
\end{tabular}
\caption{Results of a linear fit to the relation between log(variance) and log(\redd), for the four  timescales shown in Fig.~\ref{variance_vs_REdd} and the four ranges in M with four or more bins in \redd . The parameter $a$ represents the slope of the relation and parameter $b$ the log(variance) at log(\redd)$=-1$, i.e., log(variance)= $a\times\log($\redd$/0.1) +b$. For each bin in M, the slope $a$ becomes more negative when the variance is measured for shorter timescale fluctuations. }
     \label{tab:var_vs_REdd}
 \end{table}

\begin{figure}
\centering
\includegraphics[width=0.5\textwidth]{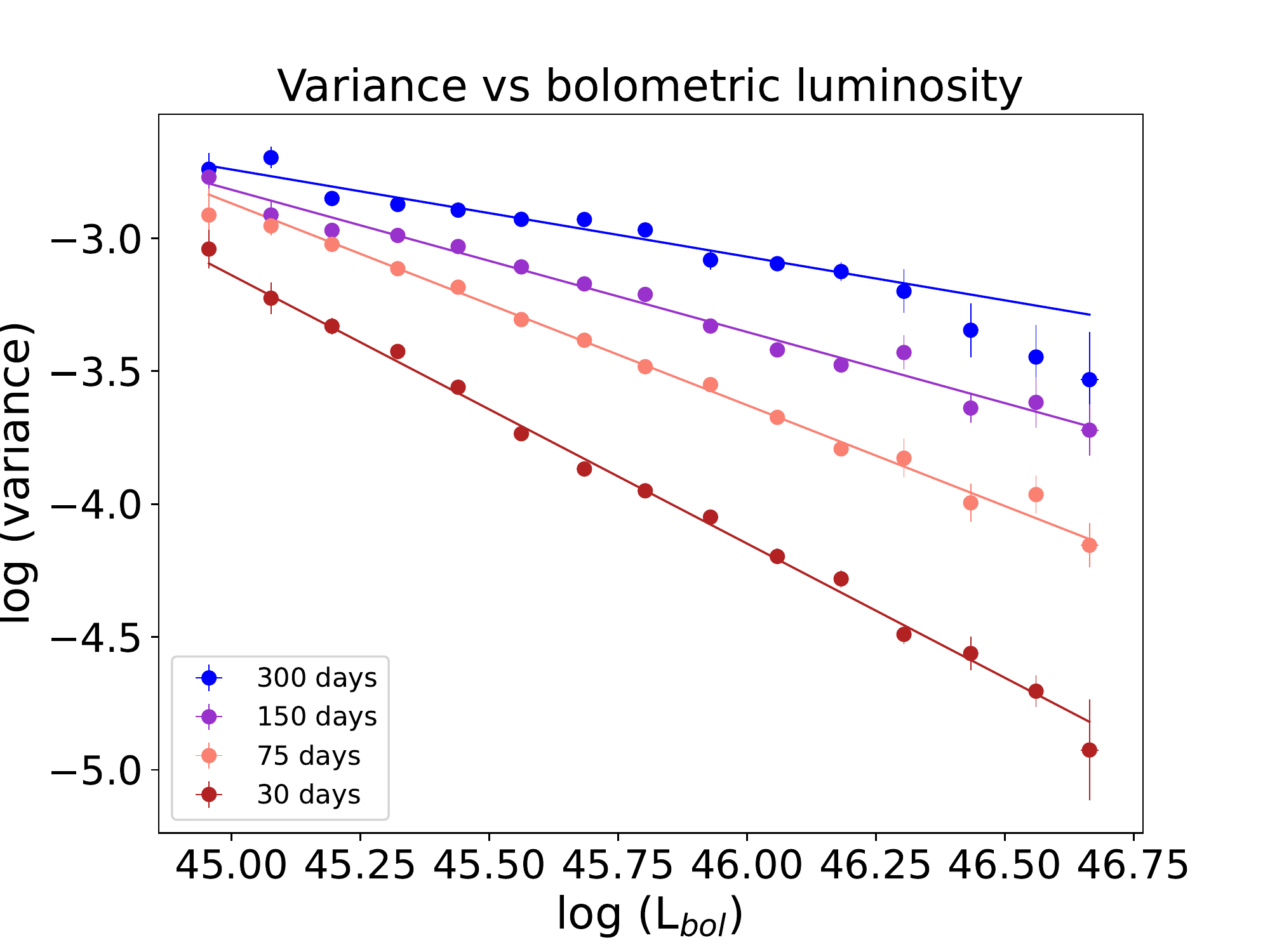}
\vspace{0.5cm}
\caption{Variance on four different variability timescales as a function of bolometric luminosity (Lbol). The markers show the median variance for different bins in Lbol for the whole range covered by our sample. The solid lines represent best-fitting linear relations to log(var) vs log(Lbol). All the error bars on the median variance of the binned data (crosses) were calculated as the root-mean-squared scatter of the medians obtained by bootstrapping using 1000 re-samples per bin in Lbol.  \label{fig:lbol}}
\end{figure}

We also computed the median variance in logarithmic bins of the bolometric luminosity, L$_{\rm bol}$, for L$_{\rm bol}$ between $10^{45}$ and $10^{46.8}$ erg s$^{-1}$, for the four timescales of variability studied. These median variances, together with the best-fitting linear models,  log(variance)= $a\times(\log$(L$_{\rm Bol})-45.8) +\ b$ are plotted in Fig.~\ref{fig:lbol}. The best-fitting values of the linear parameters $a$ and $b$ are summarized in Table \ref{tab:Lbol}.

 \begin{table}
     \centering
     \begin{tabular}{r| r r r r}
timescale[d] & \multicolumn{2}{c}{$a$} & \multicolumn{2}{c}{$b$}  \\
\hline
\hline
 300 &-0.33 &$\pm{0.03}$ &-3.01&$\pm{0.01}$\\
 150 & -0.53&$\pm{0.02}$ &-3.25 &$\pm{0.01}$\\
 75 & -0.76&$\pm{0.01}$ &-3.48 &$\pm{0.01}$\\
30 & -1.01&$\pm{0.02}$ &-3.95 &$\pm{0.01}$\\

\hline
\end{tabular}
\caption{Results of a linear fit to the relation between log(variance) and log(L$_{\rm Bol}$), for the four timescales shown in Fig.~\ref{fig:lbol}. The parameter $a$ represents the slope of the relation and parameter $b$ the  log(variance) at log(L$_{\rm Bol})=45.8$, i.e., log(variance)= $a\times(\log$(L$_{\rm Bol})-45.8) +\ b$.}
     \label{tab:Lbol}
 \end{table}

\section{Discussion}\label{sec3}
The first striking fact to notice when looking at the variance of fluctuations on different timescales as a function of black hole mass M (see Fig.~\ref{scatterplots}), is that the direction of the correlation changes depending on which timescale of variability is considered. As seen in Fig.~\ref{scatterplots} for variations on 300 days there is only a weak, positive correlation between M and variance. When only shorter timescale variations are considered, however, the relation becomes negative. The Spearman rank coefficients and related p-values for the full sample (summarized in Table \ref{spearman}) corroborate this observation, returning a weak but significant positive correlation for timescales of 300 days, and very weak correlations at 150 days and 75 days, becoming negative and more significant for 30 days fluctuations. 

It is important to note that, since both high mass and high accretion rate quasars are rare when compared to lower masses and lower accretion rates \citep[e.g.][]{Aird17, KellyMerloni11}, flux-limited samples of quasars such as this one, have a built-in anti-correlation between both parameters. This happens because the numerous quasars with both low mass and low accretion rates are too dim to be detected, allowing only the high accretion rate low-mass objects to be included. On the other hand, if the mass is high then both high and low accretion rate objects can be detected. However, objects with both high mass and high accretion rate are rare, so the high mass quasars in the sample have on average lower accretion rates than the low mass objects. This fact is a problem when studying correlations of variance with mass because the variance is known to depend on the accretion rate.  

To circumvent this problem, it is useful to split the sample into a grid based on both parameters \citep[e.g.][]{Zuo12}. The correlations between variance and mass can then be studied, for example, on sub-samples with a narrow range in \redd\ and a broad range in M. By selecting only quasars within the narrow ranges in \redd\ noted in Table \ref{spearman} the picture changes: when the dependence on \redd\ is controlled, \emph{the relation between the variance and M is always negative, becoming stronger and more significant for fluctuations on shorter timescales} e.g., as measured with shorter light curves. For all timescales, it can be seen that the negative correlation is stronger for each sub-sample than for the full sample. This is due to the fact that for the full sample the anti-correlations between \redd\ and both M and variance tend to flip the anti-correlation between variance and M. 

This effect is more easily seen in Fig.~\ref{scatterplots} where the markers represent individual variance measurements color-coded by \redd\ ranges. There is a noticeable anti-correlation between variance and M, but this is only apparent once \redd\ is considered, for example, by focusing on a fixed \redd\ (one color in this plot). For shorter variability timescales ---i.e. right plot in Fig.~\ref{scatterplots}--- the anti-correlation between variance and M is stronger. The steepening of the anti-correlation between variance and M can also be seen in the binned variances by comparing the panels in Figs.~\ref{variance_vs_mass}, which plot these relations for the four timescales of variability studied. When looking at variations on shorter timescales, for example, by using shorter light curves, the slope of the relation between variance and both M and \redd\ becomes steeper. These changes in slope are significant, as can be seen in the linear fits to these relations with parameters and errors given in Tables \ref{tab:var_vs_mass} and \ref{tab:var_vs_REdd}.

 \subsection{Implications for the power spectrum}
\begin{figure*}%
\centering
\includegraphics[scale=0.5,trim=60 0 0 0]{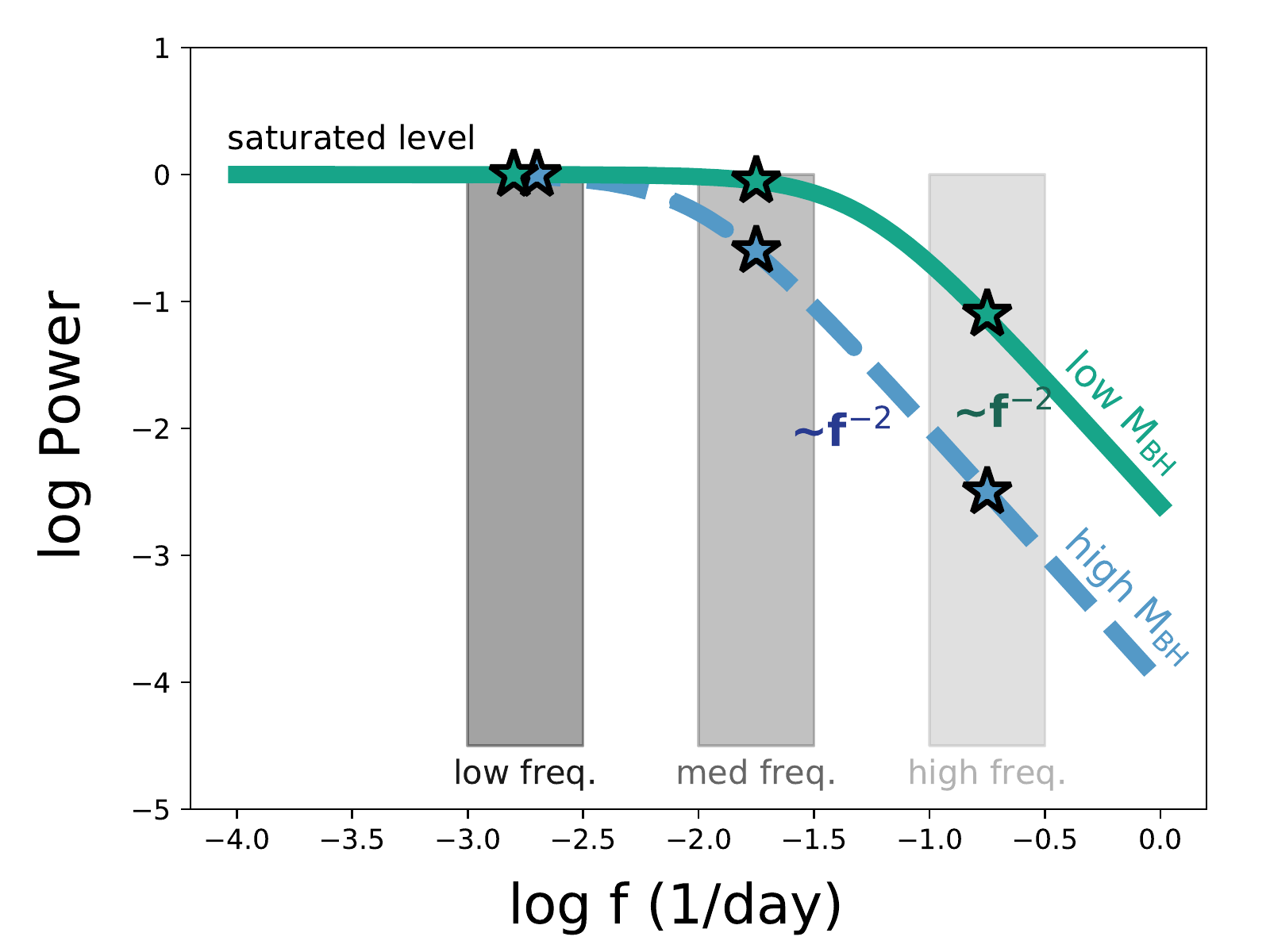}%
\includegraphics[scale=0.5,trim=0 0 0 0]{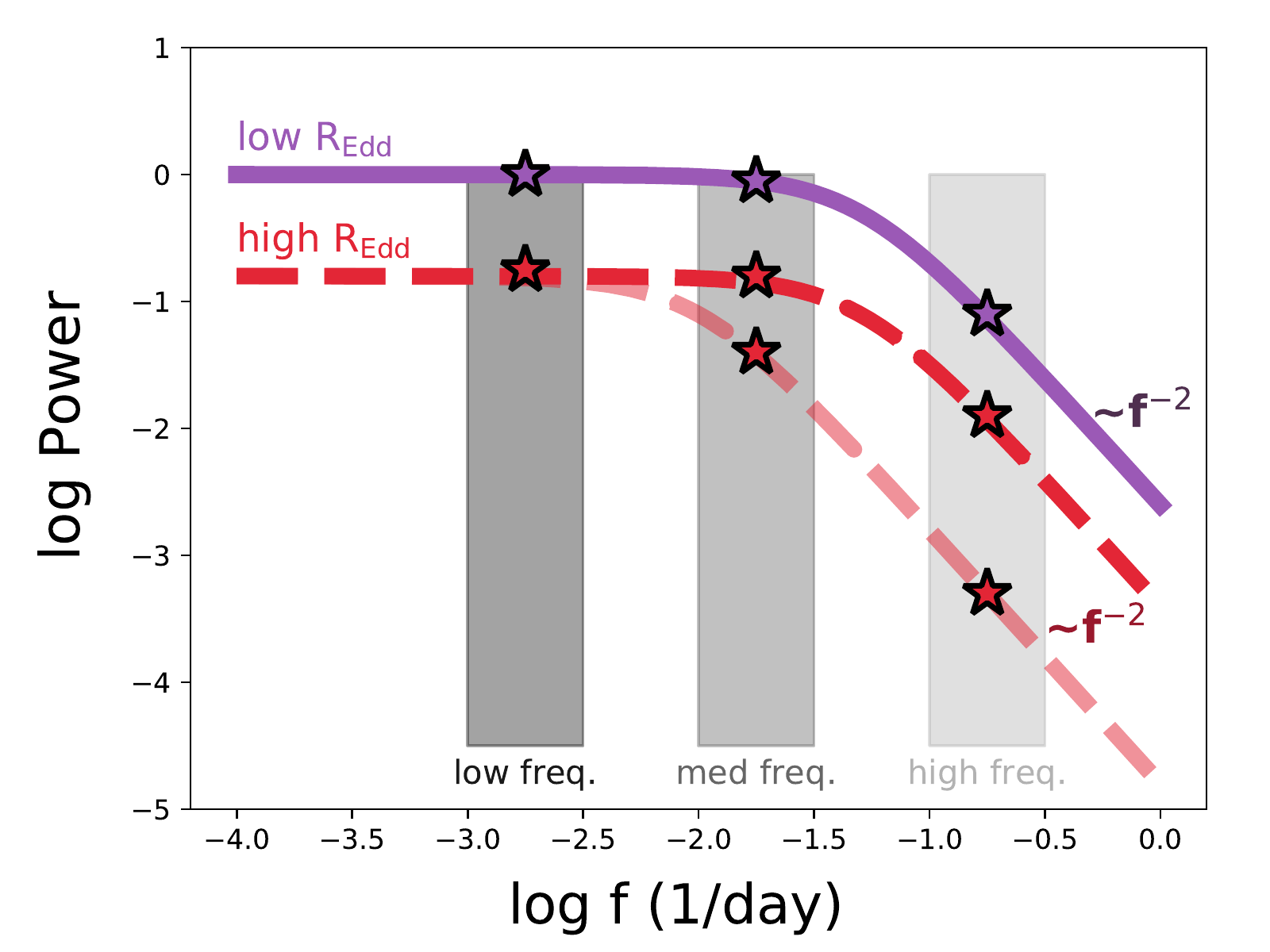}
\caption{A possible model for the dependence of the power spectrum on mass and accretion rate. Mass dependence (Left): Schematic representation of a broken power law model power spectra where the break frequency scales inversely with mass. The three shaded regions mark three timescales where the filtered variance is estimated, producing the variance levels marked by the stars. This setup produces no mass dependence of the variance at low frequencies, a strong dependence on mass at high frequencies and a weaker dependence for intermediate frequencies. Accretion rate dependence (Right): Power spectral models schematically representing different accretion rate levels for a fixed black hole mass, where the variance anti-correlates with the accretion rate. If only the normalization of the power spectra changed, the vertical distance between the low and high accretion rate power spectra would be the same in all the shaded regions (solid symbols). Instead, we observe a \emph{larger} dependence of variance with accretion rate at higher frequencies, which requires an accretion-rate dependent break timescale or high-frequency slope (symbols on the semi-transparent spectrum).}\label{PDS_mass}

\end{figure*}

Light curves can be characterized through the power spectrum, which quantifies the amount of variance found on different timescales $t$ or, equivalently, on different frequencies $f=1/t$. In the past, attempts have been made to fit quasars optical power spectra with a damped random walk model (DRW), which produces a flat, $P(f) = A $ power spectrum at low temporal frequencies $f$, steepening to $P(f) = A (f/f_b)^{-2} $ above a characteristic (break) frequency $f_b$. Even if the shape is not exactly correct, this picture is useful to interpret the variance results: first, if all quasars have essentially the same power spectrum, differing only on the break timescales, the plots in Fig. \ref{variance_vs_mass} can be qualitatively explained if the break frequency decreases with increasing mass so that, at a given variability timescale, the power spectra of larger black hole masses are probed further above the break, where the power is increasingly lower than $A$. In this sense, a positive relation between M and the characteristic timescale of DRW models fit to individual quasars was recently reported \citep{Burke21}. This scenario is schematically presented in the left panel in Fig.~\ref{PDS_mass}. In the present case, the break timescale ($T_b=1/f_b$) would be about 300 days or shorter for all the masses in the sample since the mass dependence starts to appear more strongly on shorter timescales. An exception can be perhaps the highest mass bin, where even 300 days is not long enough to reach the flat part of the power spectrum and the measured variance is still below A. In other words, for fluctuations on timescales of 300 days there is almost no mass dependence of the variance because, for most of the masses, this timescale is longer than the characteristic (break) timescale. As a result, the variance on this timescale is probing the flat part of the power spectrum, which would be the same for all masses in a given \redd\ bin. 

For the range of masses and accretion rates used here, the variance anti-correlates with mass in all accretion rate bins. The relation between the logarithms of both variance and mass appears linear in the parameter range studied and can be modeled as a linear relation as $\log $(variance)= $a\times\log {\rm M/M_{8.5}} + b$, where M$_{8.5}=10^{8.5}$M$_\odot$. The slope of these relations is small or consistent with 0 for the longest timescale fluctuations probed (300 days) and gets consistently steeper for shorter timescale variations, as can be seen comparing the panels in Fig. \ref{variance_vs_mass}, and the fitted values of $a$ in Table \ref{tab:var_vs_mass}. This behavior reconciles two apparently conflicting results: the independence of variance with mass at about one year timescales, even when controlling for different accretion rates  \citep{Wilhite08,Sanchez-Saez18}, with the mass dependence of a characteristic timescale in the variability \citep{Burke21}. The dependence of variance on mass only appears for variations on short timescales (i.e., shorter than the characteristic timescale of the power spectrum) and is small or negligible for longer timescales.

The power spectrum must depend on \redd\ as well as the M, or all the different color lines in the plots in Fig. \ref{variance_vs_mass} would fall on top of each other. The simplest interpretation would be that the normalization of the power spectrum $A$ increases with decreasing \redd. This conclusion can be reached if only the total variance of the light curves or the limiting value of the long timescale power spectrum is calculated, as opposed to this timescale-dependent approach.

An anti-correlation between power spectral normalization $A$ and \redd\ is, however, not sufficient to explain why the dependence of variance on \redd\ differs for different timescales, as evident in the different panels in Fig. \ref{variance_vs_REdd} %and in Fig. \ref{fig:A}. 
If that were the case then the power spectra of objects of the same mass and different accretion rates would simply be shifted vertically with respect to each other. By construction, the dependence of variance on \redd\ (for a fixed M) would be independent of the timescale, i.e. it would have the same slope, fixed by the overall amplitude as a function of \redd\ $A$(\redd), in all the panels in Fig. \ref{variance_vs_REdd}.  A linear fit to the relations between variance and accretion rate, as $\log$(variance)= $a\times\log$(\redd/0.1) $+\ b$ results in the best-fitting $a$ and $b$ parameters listed in Table \ref{tab:var_vs_REdd}. For each bin in mass, the slope of the relation between variance and \redd\ becomes steeper for shorter variability timescales. For the two most populated ranges in M ($8.2<$log(M)$<8.5$ and $8.5<$log(M)$<8.8$, the slopes change from  $a=-0.53\pm{0.04}$ and $a=-0.61\pm{0.06}$ at timescales of 300 days to $a=-0.95\pm{0.14}$ and $a=-0.96\pm{0.08}$ for timescales of 30 days (see Table \ref{tab:var_vs_REdd}). This small but significant steepening proves that the power spectral \emph{shape} must depend on \redd\ as well as M. This conclusion is independent of the power-spectral model considered. 
 
Returning to the bending power law model for the power spectrum, the dependence of variance on accretion rate can be achieved if either the high-frequency slope is steeper for higher accretion rates or if the break timescale $T_b$ scales not just with mass but with accretion rate as well. This possibility is exemplified in the right panel in Fig. \ref{PDS_mass}.

As seen above, the optical variance in quasars with BH masses in the range $7.5 <$ log(M) $< 9.5$ and accretion rates in the range $-2 <$ log(\redd) $< 0$ anti-correlates with both mass and accretion rate, at least for variability timescales of 150 days or shorter. On longer timescales of 300 days, the dependence on mass becomes almost negligible, and the dependence on accretion rate is reduced but still significant. Since the quasars' luminosity is the product of M and \redd, the anti-correlation of the variance with both factors results in a strong anti-correlation of variance with luminosity. The anti-correlation is especially strong for short timescales of 150 days and below where the variance has a strong anti-correlation with both mass and accretion rate. Fig.~\ref{fig:lbol} shows the filtered power on the four timescales studied, as a function of the bolometric luminosity, $L_{\rm bol}$, reported by \citet{Rakshit20}. This well known anti-correlation can have different slopes depending, for example, on the length of the light curves used, which affects the variability timescales probed when calculating the total variance. In our case, for long term fluctuations of 300 days, the relation has a best-fitting exponential slope of $-0.33\pm{0.03}$, steepening for shorter timescale fluctuations down to $-0.53\pm{0.02}$ for variations on 150 days timescales, $-0.76\pm{0.01}$ for variations on 75 days timescales, and $-1.01\pm{0.02}$ for variations on 30 days timescales. 

\section{Conclusion}

We have shown that at a given timescale, the amplitude of optical variability in quasars decreases with the black hole mass. For the first time, we show that this trend is stronger on shorter timescales (30 - 150 days in the quasar rest frame) and weaker on scales $\sim$300 days. These results reconcile previously reported weak or null dependence of variance on the black hole mass detected at long timescales with the detection of a mass-dependent break timescale in the optical light curves \citep{Burke21}. Namely, at timescales longer than the break, the correlation is weak or absent, while at shorter timescales there is a negative correlation. Such a behavior is expected in this model (See Fig.~\ref{PDS_mass}, left) and is indeed found in our work (Fig.~\ref{variance_vs_mass}). For the whole sample, the correlation between the variance and mass is weak, with a shallow trend and large scatter (Fig.~\ref{scatterplots} and Table \ref{spearman}). As we show, this scatter is largely due to the additional dependence of variance on the accretion rate and that for sub-samples covering a narrower range of accretion rates, the negative correlation between the variance and mass is stronger and correlation is tighter. The positive correlation between variance and mass that was seen on long timescales in a flux-limited sample of quasars is spurious. It is caused by the anti-correlation of the accretion rate and mass in the sample and by the well-known anti-correlation between the accretion rate and variance. 

Our analysis shows that to properly determine a link between black hole physical properties and variability, it is necessary to account not only for such properties in a well-controlled fashion but also to quantify the variability taking into account the shape of the power spectrum and {\em its} dependence on properties such as M and \redd . We have shown that this approach can explain many previous conflicting results, and yields important clues about the variability origin, as shown by the discovered dependence of the power spectrum slope at high frequencies on \redd.

It is noteworthy that the variability of quasars in the optical band does not follow the pattern observed in the X-ray domain, where the break timescale shows a linear relation with M and an inverse relation with \redd \citep{McHardy06,gonzalez12}. In the optical bands, we see a positive correlation between the break timescale and \redd, which points to a different mechanism as the driver behind these variations. Results of the continuum reverberation mapping also show that besides the intrinsic variability, the optical band includes reprocessing of a high-energy, highly variable emission, which introduces a variable signature at short time scales. Finally, as suggested by theoretical modeling \citep[e.g.][]{Kubota18}, \redd\ might control the level of X-ray reprocessing into the optical band and could explain our observed dependence of power spectrum shape on the accretion rate. 

\vspace{1cm}

{\bf Acknowledgements:} We thank the anonymous referee for helping us to improve the presentation of our results. The authors acknowledge support from the National Agency for Research and Development (ANID) grants: Millennium Science Initiative Program ICN12\_12009 (PSS,LHG), and NCN$19\_058$ (PA, PL); FONDECYT Regular 1201748 (PL); FONDECYT Postdoctorado 3200250 (PSS); Programa de Becas/Doctorado Nacional 21200718 (EL) and 21222298 (PP), and from the Max-Planck Society through a Partner Group grant (PA). 
Based on observations obtained with the Samuel Oschin Telescope 48-inch and the 60-inch Telescope at the Palomar Observatory as part of the Zwicky Transient Facility project. ZTF is supported by the National Science Foundation under Grants No. AST-1440341 and AST-2034437 and a collaboration including current partners Caltech, IPAC, the Weizmann Institute for Science, the Oskar Klein Center at Stockholm University, the University of Maryland, Deutsches Elektronen-Synchrotron and Humboldt University, the TANGO Consortium of Taiwan, the University of Wisconsin at Milwaukee, Trinity College Dublin, Lawrence Livermore National Laboratories, IN2P3, University of Warwick, Ruhr University Bochum, Northwestern University and former partners the University of Washington, Los Alamos National Laboratories, and Lawrence Berkeley National Laboratories.
Operations are conducted by COO, IPAC, and UW.\\

{\bf Data Availability Statement:} All data can be downloaded from the Zwicky Transient Facility (ZTF) Data Release 14 (DR14)\citep{Masci19}.\\

{\bf Code Availability Statement:} The Mexican Hat filter code \citep{Arevalo12} can be requested to the corresponding author.\\

\bibliographystyle{mnras}
\bibliography{bibliography} 

\begin{thebibliography}{}
\makeatletter
\relax
\def\mn@urlcharsother{\let\do\@makeother \do\$\do\&\do\#\do\^\do\_\do\%\do\~}
\def\mn@doi{\begingroup\mn@urlcharsother \@ifnextchar [ {\mn@doi@}
  {\mn@doi@[]}}
\def\mn@doi@[#1]#2{\def\@tempa{#1}\ifx\@tempa\@empty \href
  {http://dx.doi.org/#2} {doi:#2}\else \href {http://dx.doi.org/#2} {#1}\fi
  \endgroup}
\def\mn@eprint#1#2{\mn@eprint@#1:#2::\@nil}
\def\mn@eprint@arXiv#1{\href {http://arxiv.org/abs/#1} {{\tt arXiv:#1}}}
\def\mn@eprint@dblp#1{\href {http://dblp.uni-trier.de/rec/bibtex/#1.xml}
  {dblp:#1}}
\def\mn@eprint@#1:#2:#3:#4\@nil{\def\@tempa {#1}\def\@tempb {#2}\def\@tempc
  {#3}\ifx \@tempc \@empty \let \@tempc \@tempb \let \@tempb \@tempa \fi \ifx
  \@tempb \@empty \def\@tempb {arXiv}\fi \@ifundefined
  {mn@eprint@\@tempb}{\@tempb:\@tempc}{\expandafter \expandafter \csname
  mn@eprint@\@tempb\endcsname \expandafter{\@tempc}}}

\bibitem[\protect\citeauthoryear{Aird, Coil  \& Georgakakis}{Aird
  et~al.}{2017}]{Aird17}
Aird J.,  Coil A.~L.,   Georgakakis A.,  2017, \mn@doi [\mnras]
  {10.1093/mnras/stx2700}, 474, 1225–1249

\bibitem[\protect\citeauthoryear{{Angione} \& {Smith}}{{Angione} \&
  {Smith}}{1972}]{Angione72}
{Angione} R.~J.,  {Smith} H.~J.,  1972, in {Evans} D.~S.,  {Wills} D.,
  {Wills} B.~J.,  eds,  IAU Symposium Vol. 44, External Galaxies and
  Quasi-Stellar Objects. p.~171

\bibitem[\protect\citeauthoryear{{Ar{\'e}valo}, {Churazov}, {Zhuravleva},
  {Hern{\'a}ndez-Monteagudo}  \& {Revnivtsev}}{{Ar{\'e}valo}
  et~al.}{2012}]{Arevalo12}
{Ar{\'e}valo} P.,  {Churazov} E.,  {Zhuravleva} I.,  {Hern{\'a}ndez-Monteagudo}
  C.,   {Revnivtsev} M.,  2012, \mn@doi [\mnras]
  {10.1111/j.1365-2966.2012.21789.x}, \href
  {https://ui.adsabs.harvard.edu/abs/2012MNRAS.426.1793A} {426, 1793}

\bibitem[\protect\citeauthoryear{{Burke} et~al.,}{{Burke}
  et~al.}{2021}]{Burke21}
{Burke} C.~J.,  et~al., 2021, \mn@doi [Science] {10.1126/science.abg9933},
  \href {https://ui.adsabs.harvard.edu/abs/2021Sci...373..789B} {373, 789}

\bibitem[\protect\citeauthoryear{{Cristiani}, {Trentini}, {La Franca}  \&
  {Andreani}}{{Cristiani} et~al.}{1997}]{Cristiani97}
{Cristiani} S.,  {Trentini} S.,  {La Franca} F.,   {Andreani} P.,  1997, \aap,
  321, 123

\bibitem[\protect\citeauthoryear{{Dekany} et~al.,}{{Dekany}
  et~al.}{2020}]{Dekany2020}
{Dekany} R.,  et~al., 2020, \mn@doi [\pasp] {10.1088/1538-3873/ab4ca2}, \href
  {https://ui.adsabs.harvard.edu/abs/2020PASP..132c8001D} {132, 038001}

\bibitem[\protect\citeauthoryear{{Gonz{\'a}lez-Mart{\'\i}n} \&
  {Vaughan}}{{Gonz{\'a}lez-Mart{\'\i}n} \& {Vaughan}}{2012}]{gonzalez12}
{Gonz{\'a}lez-Mart{\'\i}n} O.,  {Vaughan} S.,  2012, \mn@doi [\aap]
  {10.1051/0004-6361/201219008}, \href
  {https://ui.adsabs.harvard.edu/abs/2012A&A...544A..80G} {544, A80}

\bibitem[\protect\citeauthoryear{{Hook}, {McMahon}, {Boyle}  \& {Irwin}}{{Hook}
  et~al.}{1994}]{Hook94}
{Hook} I.~M.,  {McMahon} R.~G.,  {Boyle} B.~J.,   {Irwin} M.~J.,  1994, \mn@doi
  [\mnras] {10.1093/mnras/268.2.305}, \href
  {http://adsabs.harvard.edu/abs/1994MNRAS.268..305H} {268, 305}

\bibitem[\protect\citeauthoryear{Kelly \& Merloni}{Kelly \&
  Merloni}{2011}]{KellyMerloni11}
Kelly B.,  Merloni A.,  2011, \mn@doi [Advances in Astronomy]
  {10.1155/2012/970858}, 2012

\bibitem[\protect\citeauthoryear{{Kelly}, {Bechtold}  \&
  {Siemiginowska}}{{Kelly} et~al.}{2009}]{Kelly09}
{Kelly} B.~C.,  {Bechtold} J.,   {Siemiginowska} A.,  2009, \mn@doi [\apj]
  {10.1088/0004-637X/698/1/895}, \href
  {http://adsabs.harvard.edu/abs/2009ApJ...698..895K} {698, 895}

\bibitem[\protect\citeauthoryear{{Kelly}, {Treu}, {Malkan}, {Pancoast}  \&
  {Woo}}{{Kelly} et~al.}{2013}]{Kelly13}
{Kelly} B.~C.,  {Treu} T.,  {Malkan} M.,  {Pancoast} A.,   {Woo} J.-H.,  2013,
  \mn@doi [\apj] {10.1088/0004-637X/779/2/187}, \href
  {http://adsabs.harvard.edu/abs/2013ApJ...779..187K} {779, 187}

\bibitem[\protect\citeauthoryear{{Kubota} \& {Done}}{{Kubota} \&
  {Done}}{2018}]{Kubota18}
{Kubota} A.,  {Done} C.,  2018, preprint, \href
  {http://adsabs.harvard.edu/abs/2018arXiv180400171K} {} (\mn@eprint {arXiv}
  {1804.00171})

\bibitem[\protect\citeauthoryear{{Li}, {McGreer}, {Wu}, {Fan}  \& {Yang}}{{Li}
  et~al.}{2018}]{Li18}
{Li} Z.,  {McGreer} I.~D.,  {Wu} X.-B.,  {Fan} X.,   {Yang} Q.,  2018, \mn@doi
  [\apj] {10.3847/1538-4357/aac6ce}, \href
  {http://adsabs.harvard.edu/abs/2018ApJ...861....6L} {861, 6}

\bibitem[\protect\citeauthoryear{Lu et~al.,}{Lu et~al.}{2019}]{Lu19}
Lu K.-X.,  et~al., 2019, \mn@doi [\apj] {10.3847/1538-4357/ab16e8}, 877, 23

\bibitem[\protect\citeauthoryear{{MacLeod} et~al.,}{{MacLeod}
  et~al.}{2010}]{MacLeod10}
{MacLeod} C.~L.,  et~al., 2010, \mn@doi [\apj] {10.1088/0004-637X/721/2/1014},
  \href {http://adsabs.harvard.edu/abs/2010ApJ...721.1014M} {721, 1014}

\bibitem[\protect\citeauthoryear{{Masci} et~al.,}{{Masci}
  et~al.}{2019}]{Masci19}
{Masci} F.~J.,  et~al., 2019, \mn@doi [\pasp] {10.1088/1538-3873/aae8ac}, \href
  {https://ui.adsabs.harvard.edu/abs/2019PASP..131a8003M} {131, 018003}

\bibitem[\protect\citeauthoryear{{McHardy}, {Koerding}, {Knigge}, {Uttley}  \&
  {Fender}}{{McHardy} et~al.}{2006}]{McHardy06}
{McHardy} I.~M.,  {Koerding} E.,  {Knigge} C.,  {Uttley} P.,   {Fender} R.~P.,
  2006, \mn@doi [\nat] {10.1038/nature05389}, \href
  {http://adsabs.harvard.edu/abs/2006Natur.444..730M} {444, 730}

\bibitem[\protect\citeauthoryear{{Rakshit} \& {Stalin}}{{Rakshit} \&
  {Stalin}}{2017}]{Rakshit17}
{Rakshit} S.,  {Stalin} C.~S.,  2017, \mn@doi [\apj]
  {10.3847/1538-4357/aa72f4}, \href
  {http://adsabs.harvard.edu/abs/2017ApJ...842...96R} {842, 96}

\bibitem[\protect\citeauthoryear{{Rakshit}, {Stalin}  \&
  {Kotilainen}}{{Rakshit} et~al.}{2020}]{Rakshit20}
{Rakshit} S.,  {Stalin} C.~S.,   {Kotilainen} J.,  2020, \mn@doi [\apjs]
  {10.3847/1538-4365/ab99c5}, \href
  {https://ui.adsabs.harvard.edu/abs/2020ApJS..249...17R} {249, 17}

\bibitem[\protect\citeauthoryear{{S{\'a}nchez-S{\'a}ez}, {Lira},
  {Mej{\'{\i}}a-Restrepo}, {Ho}, {Ar{\'e}valo}, {Kim}, {Cartier}  \&
  {Coppi}}{{S{\'a}nchez-S{\'a}ez} et~al.}{2018}]{Sanchez-Saez18}
{S{\'a}nchez-S{\'a}ez} P.,  {Lira} P.,  {Mej{\'{\i}}a-Restrepo} J.,  {Ho}
  L.~C.,  {Ar{\'e}valo} P.,  {Kim} M.,  {Cartier} R.,   {Coppi} P.,  2018,
  \mn@doi [\apj] {10.3847/1538-4357/aad7f9}, \href
  {http://adsabs.harvard.edu/abs/2018ApJ...864...87S} {864, 87}

\bibitem[\protect\citeauthoryear{{Simm}, {Salvato}, {Saglia}, {Ponti},
  {Lanzuisi}, {Trakhtenbrot}, {Nandra}  \& {Bender}}{{Simm}
  et~al.}{2016}]{Simm16}
{Simm} T.,  {Salvato} M.,  {Saglia} R.,  {Ponti} G.,  {Lanzuisi} G.,
  {Trakhtenbrot} B.,  {Nandra} K.,   {Bender} R.,  2016, \mn@doi [\aap]
  {10.1051/0004-6361/201527353}, 585, A129

\bibitem[\protect\citeauthoryear{{Thanjavur}, {Ivezi{\'c}}, {Allam}, {Tucker},
  {Smith}  \& {Gwyn}}{{Thanjavur} et~al.}{2021}]{Thanjuvar21}
{Thanjavur} K.,  {Ivezi{\'c}} {\v{Z}}.,  {Allam} S.~S.,  {Tucker} D.~L.,
  {Smith} J.~A.,   {Gwyn} S.,  2021, \mn@doi [\mnras] {10.1093/mnras/stab1452},
  \href {https://ui.adsabs.harvard.edu/abs/2021MNRAS.505.5941T} {505, 5941}

\bibitem[\protect\citeauthoryear{{Vanden Berk} et~al.,}{{Vanden Berk}
  et~al.}{2004}]{VandenBerk04}
{Vanden Berk} D.~E.,  et~al., 2004, \mn@doi [\apj] {10.1086/380563}, \href
  {http://adsabs.harvard.edu/abs/2004ApJ...601..692V} {601, 692}

\bibitem[\protect\citeauthoryear{{Wilhite}, {Brunner}, {Grier}, {Schneider}  \&
  {vanden Berk}}{{Wilhite} et~al.}{2008}]{Wilhite08}
{Wilhite} B.~C.,  {Brunner} R.~J.,  {Grier} C.~J.,  {Schneider} D.~P.,
  {vanden Berk} D.~E.,  2008, \mn@doi [\mnras]
  {10.1111/j.1365-2966.2007.12655.x}, \href
  {http://adsabs.harvard.edu/abs/2008MNRAS.383.1232W} {383, 1232}

\bibitem[\protect\citeauthoryear{{Wold}, {Brotherton}  \& {Shang}}{{Wold}
  et~al.}{2007}]{Wold07}
{Wold} M.,  {Brotherton} M.~S.,   {Shang} Z.,  2007, \mn@doi [\mnras]
  {10.1111/j.1365-2966.2006.11364.x}, \href
  {http://adsabs.harvard.edu/abs/2007MNRAS.375..989W} {375, 989}

\bibitem[\protect\citeauthoryear{{Zuo}, {Wu}, {Liu}  \& {Jiao}}{{Zuo}
  et~al.}{2012}]{Zuo12}
{Zuo} W.,  {Wu} X.-B.,  {Liu} Y.-Q.,   {Jiao} C.-L.,  2012, \mn@doi [\apj]
  {10.1088/0004-637X/758/2/104}, \href
  {https://ui.adsabs.harvard.edu/abs/2012ApJ...758..104Z} {758, 104}

\bibitem[\protect\citeauthoryear{{de Vries}, {Becker}, {White}  \&
  {Loomis}}{{de Vries} et~al.}{2005}]{deVries2005}
{de Vries} W.~H.,  {Becker} R.~H.,  {White} R.~L.,   {Loomis} C.,  2005,
  \mn@doi [\aj] {10.1086/427393}, \href
  {https://ui.adsabs.harvard.edu/abs/2005AJ....129..615D} {129, 615}

\makeatother
\end{thebibliography}

%%%%%%%%%%%%%%%%% APPENDICES %%%%%%%%%%%%%%%%%%%%%

\appendix

\section{Spectral considerations}\label{ap:spectral}
The light curves were obtained in the ZTF $g$ filter, which covers the wavelength range 4050--5500 \AA\ with a fairly uniform transmission \citep{Dekany2020}. For the mean redshift of the objects in this sample, this band corresponds to 2454--3333 \AA , which contains, in addition to the quasar continuum emission, broad and narrow emission lines, most notably Mg II 2900\AA\ and Fe II pseudo continuum. We note that all the variances reported are normalized by the mean flux squared, so they become dimensionless. Therefore, if the broad emission lines and pseudo-continua vary in the same proportion as the quasar continuum component, they simply act as a scaling factor for the un-normalized variance in the numerator and for the total flux squared in the denominator, so their effect is canceled, i.e. the total normalized variance is the same as that of the quasar continuum component alone. If, instead, the lines did not vary, then they would contribute to the total flux in the denominator but not to the un-normalized variance in the numerator. Therefore, the normalized variance would be suppressed by a factor $1/(1+f)^2$, where $f$ is the ratio between the luminosity of the lines and the luminosity of the quasar continuum in the $g$ band. To estimate the maximum suppression of the variance that we can expect, i.e. in the case when the additional components do not vary, we retrieved the Mg II line luminosities and the luminosity at 3000 \AA\ in the restframe, reported by \citet{Rakshit20}. We crudely estimate the quasar continuum flux in the observed $g$ band by taking the luminosity at 3000\AA, reported in erg s$^{-1}$ divide them by 3000\AA\ and multiply them by the band width of the $g$ band in the restframe, i.e. 879\AA . The bulk of the objects in the sample have a ratio of Mg II line luminosity to the total luminosity in the observed $g$ band between $f=3\%$ and $f=15\%$ for the broad component and between $f=0.03\%$ and $f=1\%$ for the narrow component. From here, we conclude that the effect of any non-variable Mg II emission lines is subdominant and cannot explain the changes by factors of a few or more in the measured variance as a function of $M$ or \redd.

We also note that since all the quasars studied have $0.6<z<0.7$, the spectral region contained in the $g$ band is approximately the same for all of them, and therefore the correlations found are not produced by spectral features moving in or out of the spectral range of the filter. 

\begin{figure*}
\centering
\includegraphics[width=0.49\textwidth,trim= 0 0 0 0]{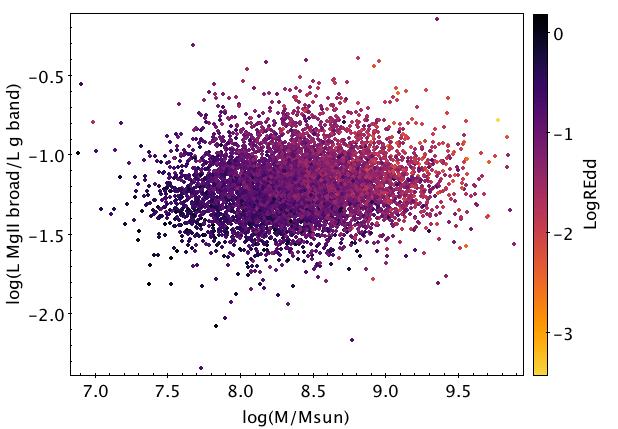}
\includegraphics[width=0.49\textwidth,trim= 0 0 0 0]{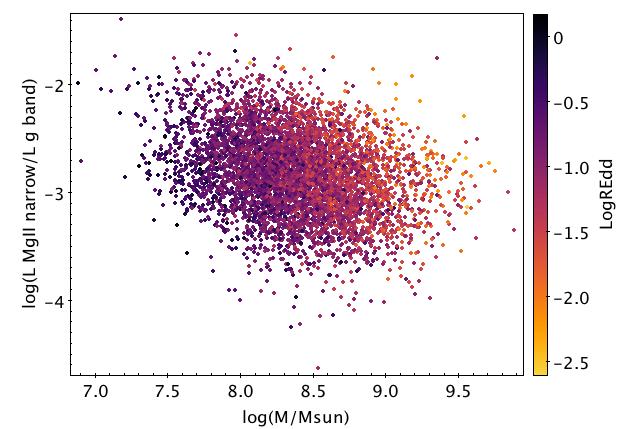}
\caption{Estimates of the ratio $f$ between MgII (2900\AA ) luminosity and the continuum luminosity in the observed $g$ band,  based on the luminosities reported by \citet{Rakshit20} for the broad (left) and narrow (right) line components. If the line luminosities were completely constant, they would lead to a maximum suppression of the normalized variance by a factor of $1/(1+f)^2$.  }
\end{figure*}

\section{Subtraction of the variance produced by observational noise}\label{ap:noise}
For each point in the light curve, we added a Gaussian deviate with $\sigma$ equal to the error on the flux of the corresponding epoch so that the variance of the simulated light curve, $\sigma^2_{\rm sim}$, would be equal to the intrinsic variance $\sigma^2_{\rm intrinsic}$, plus the observational noise variance, $\sigma^2_{\rm noise}$, plus the variance of the noise estimated from the error bars,$\sigma^2_{\rm noise, estimate}$. Specifically, the variance of the light curve with additional noise can be calculated as 
\begin{equation}
    \sigma^2_{\rm sim} =\ <lc^2>-<lc>^2\ =\ <(s+e+n)^2>-<s+n+e>^2
\end{equation} where $lc$ is the light curve with additional noise, $s$ is the intrinsic variability signal, $e$ are the deviations of each flux measurement due to observation error, and $n$ are the flux deviates added to simulate additional noise. Assuming that the intrinsic signal, the sign of the observational error, and the additional noise are uncorrelated and that the last two are distributed around zero, the cross terms and the expectation value of $n$ and $e$ can be canceled, which results in 
\begin{eqnarray}
\sigma^2_{\rm sim}&=&<s^2>+<e^2>+<n^2> - <s>^2\\
 &=& <s^2>-<s>^2+<e^2>+<n^2>,
\end{eqnarray}
i.e., the intrinsic variance, $\sigma^2_{\rm intrinsic}$ plus the variance of the noise, $<e^2>=\sigma^2_{\rm noise}$, and the variance of the additional noise, $<n^2>=\sigma^2_{\rm noise, estimate}$. In our case, we estimated the variance of the original and simulated light curves using the Mexican Hat filter to include in the noise estimate any possible biases that this method can have on the measured variances. 

Taking the difference between the variance of the simulated light curve and the original light curve results in an estimate of the variance of the noise, i.e., ($\sigma^2_{\rm sim})-(\sigma^2_{\rm original})=(\sigma^2_{\rm intrinsic}+\sigma^2_{\rm noise}+\sigma^2_{\rm noise, estimate}) - (\sigma^2_{\rm intrinsic}+\sigma^2_{\rm noise}) = \sigma^2_{\rm noise, estimate}$

If the error bars on the flux are a good representation of the observational noise then, on average, $\sigma^2_{\rm noise} = \sigma^2_{\rm noise, estimate}$ and this estimate can be subtracted from the variance of the original light curve to obtain the intrinsic variance. We repeated this procedure 10 times per light curve to obtain an average noise estimate which was then subtracted from the variance of the original light curves. The variances were always estimated with the Mexican Hat filtering described above, so the whole process was repeated for each variability timescale studied. 

\begin{figure*}%band magnitude
 \centering
\includegraphics[width=0.5\textwidth]{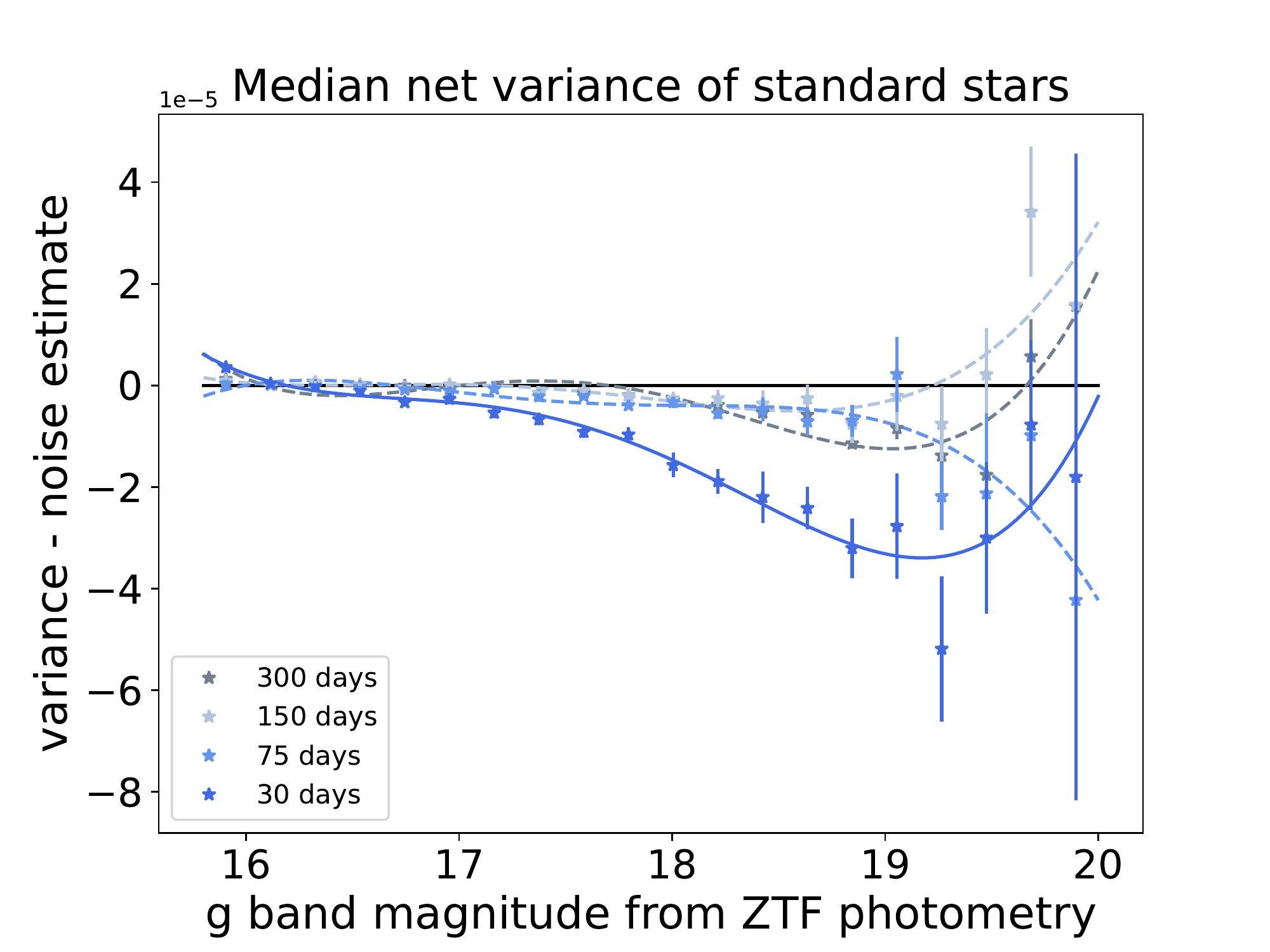}%
\includegraphics[width=0.5\textwidth]{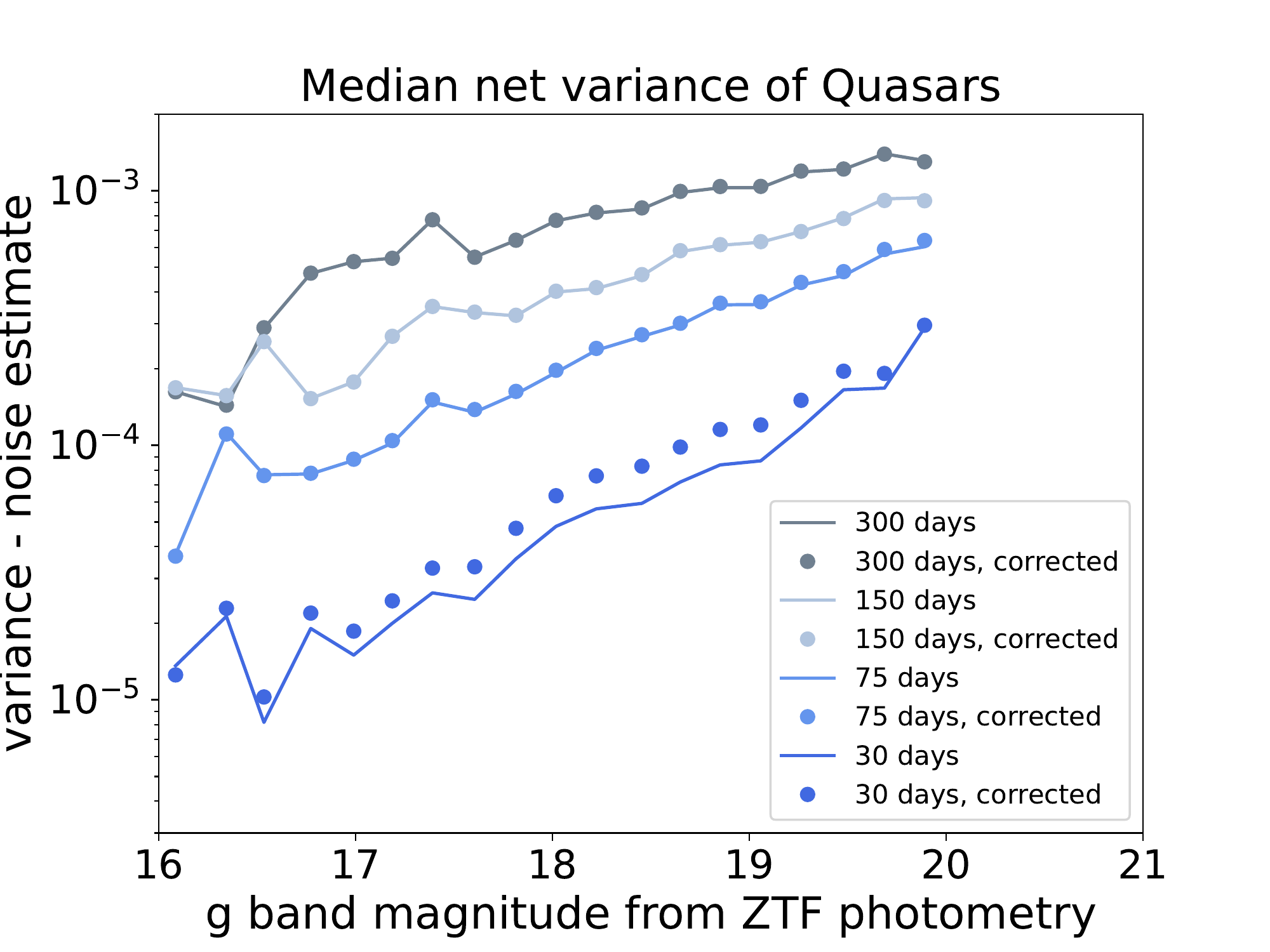}

\vspace{0.5cm}
\caption{Left: Net variance (total variance - noise estimate) of the non-variable stars as a function of ZTF $g$ magnitude, filtered on four different variability timescales as shown in the legend. Markers represent the median net variance for each bin in magnitude, and lines are polynomial fits. All the error bars on the median variance of the binned data (star markers) were calculated as the root-mean-squared scatter of the medians obtained by bootstrapping using 1000 re-samples per magnitude bin. The expectation for the net variance of these non-variable stars is zero. The fact that negative values are obtained shows that the noise is overestimated, and that this is a larger problem for dimmer objects and shorter timescales. Right: Median net variances of the quasars, before correction in solid lines and after correction in solid markers. As the quasars are intrinsically variable, all the median variances are positive and can therefore be plotted with a logarithmic $y$-axis. The overestimation of the noise, evident in the plot on the left, is corrected for by adding the ``missing variance" needed to take the non-variable stars to zero variance to the variance of each quasar, as a function of its magnitude, and is only significant for the median variance on the shortest timescale. We note that the good correlation between apparent magnitude and variance in the right plot is a by-product of the good correlation between variance and bolometric luminosity (e.g. Fig.\ref{fig:lbol}) and the fact that all the sample objects are at about the same distance from the observer. Such a tight correlation might not appear in a more heterogeneous sample.\label{fig:stars}
}
\end{figure*}

We checked the reliability of the observational noise estimates with a sample of non-variable stars observed with ZTF. The expectation is that their net variances should distribute symmetrically around 0. We selected and retrieved the ZTF light curves of 2554 stars from the SDSS Stripe 82 Standard Star catalog of \citet{Thanjuvar21} with $16< g< 20.5$, tha had at least 10 SDSS observations and a small dispersion in g band magnitude (g band rms < 0.02 as reported in the \citealt{Thanjuvar21} catalog). After applying the same cleaning and selection criteria used for the quasar light curves, we obtained 2304 valid light curves. We estimated the variance on the four different timescales using the same Mexican Hat filtering applied on the quasar light curves and estimated the variance due to observational noise as described above. The net (i.e., total - estimated noise) variances for each variability timescale are plotted in the left panel in Fig.~\ref{fig:stars}; the markers show the median net variance for different bins in $g$ band magnitude. The negative net variances show that our noise estimate based on the flux error bars slightly overestimates the observational noise. The polynomial fits to each data set (dashed lines) were used to interpolate the expected deviations in the noise estimates as a function of magnitude. This estimate was used to correct all the quasar net variances. The right panel in Fig.~\ref{fig:stars} shows the median quasar variance for different $g$ band magnitude bins and for the four variability timescales studied. After applying the correction on the noise estimate, the median variance shifted from the values of the solid lines to those shown in the markers. The correction is only significant for the shortest timescale of variability, where it amounts to about 10-20\% of the median variance. These net and corrected variances are our estimates of the intrinsic variance of each quasar used throughout this work.

\bsp	
\label{lastpage}
\end{document}